# Symphony: Expressive Secure Multiparty Computation with Coordination

Ian Sweet[a], David Darais[b], David Heath[c], William Harris[b], Ryan Estes[d], and Michael Hicks[a,e]

a University of Maryland, USA
b Galois, Inc., USA
c Georgia Institute of Technology, USA
d University of Vermont, USA
e Amazon, USA

**Abstract**

**Context** Secure Multiparty Computation (MPC) refers to a family of cryptographic techniques where mutually untrusting parties may compute functions of their private inputs while revealing only the function output.

**Inquiry** It can be hard to program MPCs correctly and efficiently using existing languages and frameworks, especially when they require coordinating disparate computational roles. How can we make this easier?

**Approach** We present Symphony, a new functional programming language for MPCs among two or more parties. Symphony starts from the single-instruction, multiple-data (SIMD) semantics of prior MPC languages, in which each party carries out symmetric responsibilities, and generalizes it using constructs that can coordinate many parties. Symphony introduces *first-class shares* and *first-class party sets* to provide unmatched language-level expressive power with high efficiency.

**Knowledge** Developing a core formal language called $\lambda$-Symphony, we prove that the intuitive, generalized SIMD view of a program coincides with its actual distributed semantics. Thus the programmer can reason about her programs by reading them from top to bottom, even though in reality the program runs in a coordinated fashion, distributed across many machines. We implemented a prototype interpreter for Symphony leveraging multiple cryptographic backends. With it we wrote a variety of MP programs, finding that Symphony can express optimized protocols that other languages cannot, and that in general Symphony programs operate efficiently.

**Grounding** In addition to developing the formal proofs, the prototype implementation, and the MPC program case studies, we measured the performance of Symphony's implementation on several benchmark programs and found it had comparable performance Obliv-C, a state-of-the-art two-party MPC framework for C, when running the same programs. We also measured Symphony's performance on an optimized *secure shuffle* protocol based on a coordination pattern that no prior language can express, and found it has far superior performance to the standard alternative.

**Importance** Programming MPCs is in increasing demand, with a proliferation of languages and frameworks. This work lowers the bar for programmers wanting to write efficient, coordinated MPCs that they can reason about and understand. The work applies to developers and cryptographers wanting to design new applications and protocols, which they are able to do at the language level, above the cryptographic details. The $\lambda$-Symphony formalization of Symphony, and the proofs about it, are also surprisingly simple, and can be a basis for follow-on formalization work in MPC and distributed programming. All code and artifacts are available, open-source.

## 1 Introduction

Secure Multiparty Computation (MPC) is a subfield of cryptography that allows mutually untrusting parties to compute arbitrary functions of their private inputs while revealing only the function output. That is, MPC allows parties to run programs *under encryption*. MPC technology today is hundreds of millions of times faster than technology from a mere 20 years ago, meaning that many more computations are now feasible. Modern frameworks are also increasingly convenient, offering programmers a familiar language in which to express their MPC programs [36, 2, 14, 39, 26, 32, 5].

Unfortunately, while these frameworks allow the programmer to express many MPCs, they fall short when confronted with problems that require non-trivial *coordination*. Overwhelmingly, MPC frameworks take the default view that all parties perform the same synchronized activity, in the style of single-instruction multiple-data (SIMD). This simplified view is not always appropriate, and in many scenarios it is useful for parties to execute *different* computations. For example, suppose we wish to implement a round-based card game where $N$ players use MPC to jointly shuffle and deal the cards. By using MPC, the parties ensure that the deck and the players' individual hands remain secret. At the same time, each party might choose to play cards according to her own strategy, and so each party might carry out different actions, perhaps by interacting with a user via I/O. Moreover, a party might drop out of the computation altogether once eliminated from the game. As another example, suppose that a very large number of parties wish to provide inputs to a privacy-preserving computation. In such situations, it is pragmatic for the parties to elect a small committee to carry out the computation on their behalf. Doing this can greatly aid efficiency, since the performance of most MPC primitives degrades as the number of parties grows.

Most existing MPC frameworks offer no coordination features; non-synchrony is handled by ad hoc mechanisms, or not at all. Ad hoc mechanisms can lead to programming mistakes, and these mistakes can result in (potentially nondeterministic) hangs or wrong answers. The language Wysteria [28] provides coordination support, but lacks *expressiveness* and *ergonomics*. For example, individual parties may not delegate computations to other parties, and MPCs must be expressed in a rigid sublanguage that makes interleaving encrypted and plaintext computation unnecessarily awkward.

**Our Approach** We present SYMPHONY, an expressive MPC language that provides appropriate tools for coordinating MPCs with large numbers of parties. SYMPHONY's most interesting language features are as follows:

- *Scoped parallel expression blocks*, or *par* blocks, allow the programmer to easily control which parties execute which code. The programmer annotates a *par* block with the set of parties that should enter the block. The language ensures that parties executing within the block agree on the logical values of local variables, so there is no risk of deadlocks or undefined behavior. Crucially, in the style of choreographic programming languages [8, 9, 10, 25], we prove that the developer can reason about her program as if it runs single threaded.

- *Party sets* are *first-class* values in the language: they can be put together and computed on at run-time, and used to annotate *par* blocks, thereby supporting highly expressive, dynamically determined coordination operations.
- *First-class shares* model multiparty-encrypted values in a way that allow the programmer to freely *delegate* computation and *reshare* encrypted values from one party set to another, and to *reactively* mix MPC operations with cleartext operations.

These features have previously appeared, in part, in other MPC frameworks: *par* blocks and party sets are inspired by Wysteria, and first-class shares are inspired by the programming style of EMP [36] and Obliv-C [39]. Symphony is the first MPC language to carefully *combine* these features, and to *generalize* them, e.g., by allowing re-sharing and delegating already encrypted values to a different set of parties, and freely constructing and *computing* on party sets to refine coordinating actions. This combination of features is crucial for writing protocols in which parties may freely enter, leave, and shift the locus of cryptographic computations.

**Contributions** Our specific contributions are as follows.

- We motivate the need for Symphony, describe its key features using examples, and compare it to the state of the art (Sections 2 and 3).
- We present a core formalism for Symphony, called $\lambda$-Symphony—syntax (Section 4.1), *single-threaded* semantics (Section 4.2), and *distributed* semantics (Section 5). We prove that the single-threaded semantics faithfully represents the distributed semantics, and thus can be used as the basis for reasoning about a Symphony program's behavior (Section 6).
- We describe our prototype implementation for Symphony, which leverages EMP [36] and MOTION [5] as its cryptographic backends (Section 7).
- We discuss 16 programs implemented in Symphony, showing that it provides *ergnomic* and *expressiveness* benefits compared to prior languages (Section 8.1).
- We show that despite its expressiveness, Symphony enjoys good performance. On a set of kernel benchmarks running on a simulated LAN, Symphony takes a mean 1.15× the time taken by Obliv-C (Section 8.2).

Symphony is publicly available at https://github.com/plum-umd/symphony-lang.

## 2 Background and Related Work

This section presents some background on related MPC programming frameworks, with detailed consideration of the problem of *coordination*.

### 2.1 Prior Frameworks, Broadly

Table 1 characterizes existing, prominent MPC frameworks according to features of the underlying MPC protocol (columns 2–5) and the ways computations are programmed (columns 6–7). As a representative example, here is "millionaires" in Obliv-C, in which two parties, A and B, wish to learn who is richer without revealing their total wealth:

**Table 1** Comparing MPC frameworks. **Proto** indicates whether the underlying cryptography mechanism uses *garbled circuits* (GC), *linear secret sharing* (LSS), or a custom *hybrid* (Hy) protocol that leverages both; # indicates the number of parties involved in a secure computation; **TM** indicates the *threat model* the protocol addresses, either *semi honest* (S) or *malicious* (M), or both; **Dom** indicates whether the framework supports MPC over boolean (B) circuits, arithmetic (A) circuits, or both; **Type** indicates if the MPC framework is a custom language (DSL), language extension (Ext), or library (Lib); and **Features** indicates the extent to which the framework supports *reactive* computations (R), *delegation* from IO parties to compute parties (D), and *synchronized randomness* ($) among multiple parties.

| | **Proto** | **#** | **TM** | **Dom** | **Type** | **Features** | | |
|---|---|---|---|---|---|---|---|---|
| | | | | | | R | D | $ |
| EMP-toolkit [36] | GC | 2 | S,M | B | Lib (C++) | ● | ○ | ◐ |
| Obliv-C [39] | GC | 2 | S | B | Ext (C) | ● | ○ | ● |
| ObliVM [24] | GC | 2 | S | B | DSL | ● | ○ | ● |
| TinyGarble [33] | GC | 2 | S | B | Ext (Verilog) | ○ | ○ | ○ |
| Wysteria [28] | LSS | 2+ | S | B | DSL | ● | ○ | ◐ |
| ABY [12] | GC,LSS | 2 | S | A,B | Lib (C++) | ○ | ○ | ◐ |
| MOTION [5] | GC,LSS | 2+ | S | A,B | Lib (C++) | ○ | ○ | ◐ |
| SCALE-MAMBA [2] | Hy | 2+ | S,M | A | DSL | ● | ○ | ● |
| Sharemind [27] | Hy | 3 | S | A | DSL | ● | ● | ○ |
| MPyC [32] | LSS | 3+ | S | A | Lib (Python) | ● | ○ | ● |
| PICCO [40] | Hy | 3+ | S | A | Ext (C) | ○ | ● | ○ |
| Frigate [26] | GC | 2 | S,M | B | DSL | ◐ | ○ | ◐ |
| CBMC-GC [14] | GC | 2 | S,M | B | Ext (C) | ○ | ○ | ○ |
| Viaduct [1] | GC,LSS | 2 | S | A,B | DSL | ○ | ○ | ◐ |
| Symphony | GC,LSS | 2+ | S | A,B | DSL | ● | ● | ● |

```
1 void millionaire(void* args) {
2   protocolIO* io = args;
3   obliv int v1 = feedOblivInt(io->mywealth, A);
4   obliv int v2 = feedOblivInt(io->mywealth, B);
5   obliv bool ge = v1 >= v2;
6   revealOblivBool(&io->cmp, ge, 0);
7 }
```

Both parties run this program, SIMD-style. Variables `v1` and `v2` are encrypted, as per the `obliv` keyword. Function `feedOblivInt` sets the initial values of these variables. On party A, line 3 encrypts the input stored in A's copy of `io->mywealth`, while line 4 does likewise for B. Internally, the function will send its encrypted input to the other party, which synchronously receives it; i.e., line 3 sends from A to B and line 4 from B to A. Line 5 computes on these encrypted values (at both parties), with the result itself being encrypted. Finally, `revealOblivBool` coordinates among the two parties to decrypt the result; it is stored in each party's `&io->cmp`.

Under the hood, these frameworks use different MPC protocols (e.g., 2-party *garbled circuits* [37], $N$-party *linear secret shares* [16, 4], or *hybrid* [11] protocols), expose various primitives operations (*boolean*, *arithmetic* or both), and consider different sorts of adversary (a "semi-honest" protocol follower, or a "malicious" protocol subverter).

Sometimes influenced by under-the-hood features, the above-the-hood programmability features are also different; Table 1 lists three: *reactive MPC* (R), *delegation* (D), and *synchronized randomness* ($). Reactive MPCs are those that invoke the underlying MPC protocol multiple times, where later invocations can leverage results revealed by earlier ones. For example, Obliv-C allows reactive MPC, so we could extend our millionaires example above to use the revealed `&io->cmp` in follow-on computation involving other inputs from `A` and `B`. Reactive MPC is useful for optimization (e.g., joint median [22] and private set intersection [20]) and interactivity (e.g., multi-round card games where the deck is encrypted).

*Delegation* refers to the ability to specify *IO parties* that provide inputs and receive outputs, but do not participate in the secure computation directly. Delegation is imporant for scalability—an MPC protocol that is intractable to run on 100 parties may be tractable to run by delegating to a committee of 3. Modern MPC protocols often provide the cryptographic support needed for delegation (by additively sharing inputs and outputs) but lack the requisite programming support to coordinate non-compute parties, which is not a good match for the basic SIMD model (see Section 2.2).

Finally, *synchronized randomness* refers to multiple parties being able to efficiently access the same stream of random numbers; this can be extremely useful in conjunction with delegation to optimize reactive MPCs [19]. Many languages that don't support it could be extended to do so, denoted with a half-circle in the table. For the others: CBMC-GC and TinyGarble compile to standard Boolean circuits, which do not have a gate for uniform bit generation; PICCO and Sharemind are based on threshold MPC protocols that require only a subset of the declared parties to execute, and synchronizing among a dynamic subset efficiently would be nontrivial.

### 2.2 Coordination

A SIMD computation model means it's easiest to specify parties that do the same thing. But what if we want them to *coordinate* parties that should behave differently, even based on dynamically determined information?

**Pitfalls** Delegation and reactive MPC are important coordinating mechanisms, but are not entirely sufficient. Some frameworks provide low-level mechanisms for the programmer to specify party-specific actions, but these are easy to misuse. To illustrate, here's an Obliv-C program in which `<code>` is executed only at party $X$:

```
if (ocCurrentParty() == X) { <code> }
```

Other Obliv-C constructs also take party identifiers to localize execution, e.g., `readInt` reads from local storage on the identified party. Using such constructs requires care. Suppose A wishes to share the encrypted result of computing `f` on its input `a`. The following code to do so contains a bug:

```
1 int a;
2 readInt(&a, "input.txt", A);
3 obliv int share;
4 if (ocCurrentParty() == A) {
5   share = feedOblivInt(f(a), A);
6 }
7 ... // proceed with secure computation on f(a)
```

Due to the conditional on line 4, the share on line 5 will trigger a `send` on A but no corresponding `recv` on B, causing A to block forever awaiting B's response. We fix the issue by dropping the conditional on line 4, so `feedOblivInt` is called at both parties.

```
1 int a;
2 readInt(&a, ..., A);
3 obliv int share = feedOblivInt(f(a), A);
```

The code no longer hangs on A, but now there is another problem: undefined behavior. The call to `readInt` on line 2 initializes `a` on A but not on B. The call to `f(a)` causes both A and B to read the value contained in `a`, causing undefined behavior on B. The solution is to provide a dummy value for `a` when executing on B.[1]

```
1 int a;
2 readInt(&a, ..., A);
3 int fa = ocCurrentParty() == A ? f(a) : 0;
4 obliv int share = feedOblivInt(fa, A);
```

There is still a risk: `f` must not perform any communication among parties, otherwise we will experience another coordination error like the one in the first example.

**Three (or more) party coordination** Coordination gets even more unwieldy when writing MPCs for $N \geq 2$ parties. In general, we might have dozens or more parties, with interactions between overlapping sets of parties. Each party's role may shift over time, possibly dependent on prior computations. These complexities are perhaps the reason that $N$-party languages like PICCO [40], Frigate [26], and SCALE-MAMBA [2] do not even provide low-level coordination mechanisms.

Wysteria [28, 30] is unique in providing higher-level support for safe $N$-party coordination. Here's the integer sharing example from Section 2.2 in Wysteria.

```
1 let a =par({A})= read() in
2 let fa =par({A})= f a in
3 let b =par({B})= 0 in (* Dummy *)
4 let inp = (wire {A} fa) ++ (wire {B} b) in
5 let share =sec({A,B})= makesh inp[A] in
```

While all parties locally execute the same program, the `let x =par(P)= e in` ... annotation indicates that only parties in `P` should execute `e`; the rest will skip it. Party-annotated expressions prevent coordination errors. For example, variable `a`, created by A on line 1, may be accessed only when A alone is in scope, as on line 2. Wysteria's type system will reject programs that try to do otherwise. When variables are visible to more than one party, Wysteria's design ensures that all parties agree on their contents.

[1] We could have instead ensured that `a` is initialized to a dummy value on B. This works, but when dealing with compound types (e.g. an array of integers) it requires allocating memory on B for all of A's input and initializing the memory with dummy values.

Wysteria also provides an annotation to specify an MPC: `let x = sec(P) e in` … says that the parties in `P` should jointly execute `e` as an MPC. When reached at run-time, the Wysteria interpreter translates `e` to a circuit and executes it with the GMW protocol, revealing the result to all parties. Inputs are specified as *wire bundles*, created via syntax `wire P e`; these are essentially maps from parties to values, with the mapped-to value visible only to the keying party. Parties package their inputs in a bundle (as on line 4) which can then be accessed within `sec` expressions via array-index notation (line 5, which gets A's input). Rather than immediately reveal the final result, programmers can mark values as shares via `makesh` (line 5); these shares can be referenced again in a later `sec` expression (among the same parties) to continue performing MPC over them (as we will see later). Once again, party annotations prevent accessing non-existent values. Moreover, the use of `sec` ensures all parties will synchronize at an MPC, avoiding deadlocks/hangs.

While better than other frameworks, Wysteria's coordination support has limitations. Firstly, while Wysteria party sets are values, operations on them are so limited that the sets can always be resolved statically when given a whole program. (See Appendix A.3 for more information.) Coordinating among parties chosen dynamically is important for security (e.g., choosing random parties to prevent cheating) and interaction (e.g., the set of players in a card game). Additionally, Wysteria does not draw a distinction between compute parties and IO parties, which precludes support for delegation even among static party sets. Wysteria shares cannot be re-shared to different parties without decrypting them.

## 3 SYMPHONY: Expressive, Coordinated MPC

SYMPHONY is a domain-specific programming language for expressing MPCs of $N \geq 2$ parties by generalizing the standard SIMD-style view. It takes inspiration from Wysteria's approach to the coordination problem, providing similar safety guarantees, but with significant enhancements to the language's expressiveness and efficiency. This section introduces SYMPHONY through examples; the next few sections develop it formally and prove its safety benefits.

### 3.1 Basics

SYMPHONY is a dynamically typed functional programming language with support for integers, pairs, variant (sum) types, lists, let-binding, pattern matching, (recursive) higher-order functions, and (mutable) references and arrays.

Recall the millionaires example implemented in Obliv-C and shown in Section 2.1. An analogous implementation in SYMPHONY is as follows:

```
1 principal A B
2 def main () = par {A,B}
3   let a = par {A} read int from "input.txt" in
4   let b = par {B} read int from "input.txt" in
5   let v1 = share [gmw, int : {A} -> {A,B}] a in
6   let v2 = share [gmw, int : {B} -> {A,B}] b in
7   let ge = v1 >= v2 in
8   reveal [gmw, bool : {A,B} -> {A,B}] ge
```

As is standard, all parties run the same program (starting at `main`).

**Par blocks** SYMPHONY uses `par` blocks to permit some, but not all, parties to execute a part of the program. For example, when `par {A}` ... is reached on line 3, only the listed party A in the set executes the subsequent code .... In this case, party A reads an integer from the file `input.txt` on its local filesystem. Parties not in a `par`'s set skip the code block, returning an *opaque value* ★ which will cause an error if computed upon (since no real value is available at that party).

**Located data** As SYMPHONY programs are fundamentally distributed, data is *located* at particular parties' hosts. An important invariant is that when different hosts access the same program variable they will see the same logical value, or else the whole program will fail because at least one host sees a ★. (SYMPHONY also provides a *bundle* abstraction like Wysteria's to allow different parties to use the same variable to hold different values.) As a result, parties are naturally coordinated, avoiding wrong results, deadlocks, and hangs like those mentioned in Section 2.2.

**Shares** Encrypted values in SYMPHONY are called *shares*, which we can think of as secret shares among a particular set of parties. (SYMPHONY implements both GMW and Yao back ends.) We use `share`$[\phi, \tau : P \rightarrow Q]\ v$ to take value $v$ now at $P$ and secret-share it among parties $Q$, where $\phi$ is the MPC protocol (e.g., `gmw`) and $\tau$ is the type of the share's contents. Oftentimes $P$ is a single party and $Q$ is a set. For example, line 5 secret-shares A's value `a` with both A and B, storing the share in `v1`. This maintains the location invariant: parties A and B agree on the *logical* value of `v1`—it is a share of the same value among those two parties. The `share` operation requires all parties $P \cup Q$ to be currently executing (ensured by the `par` on line 2) so that $P$ can transmit to each party in $Q$ its share and know they are ready to receive it. Each local executing party keeps track of the intersection of the sets of in-scope `par` blocks in order to perform this check. Shares can be operated on like normal values as long the sharing parties carry out the operation. For example, line 7 compares the shared `int`s and stores the result in `ge` (also a share), whose logical value both A and B will agree on.

**Revelation** While `share` converts its argument at $P$ to shares at $Q$, `reveal` converts shares at $P$ to plaintext at $Q$. Line 8 converts the share `ge` at A and B to a plaintext value sent to those same hosts. As with `share`, all parties in $P \cup Q$ must be executing the block or else induce an error.

**Reasoning** Although actual execution of the program occurs on multiple machines communicating with each other, Symphony allows programmers to reason about their code as if it was executed line-by-line on a single machine, which we call the "single threaded interpretation." We prove this is the case via three theorems in Section 6, the proofs for which rely on key invariants on located data and shares. The first theorem states that a program will terminate in the single threaded interpretation if-and-only-if it terminates (implying no deadlock) in the actual execution. This is useful for debugging, e.g., if the program terminates with synthetic input in a single-threaded simulation (which our implementation provides), and termination is not dependent on input data, then the actual execution of the distributed program on secret input is guaranteed to both terminate and never deadlock. The second and third theorems establish a similar correspondence but for executions that are halted due to a runtime error detected by Symphony—e.g., due to a party accessing a variable containing ★ or executing a `share` in which parties are missing—essentially showing that the single threaded execution will detect such errors if-and-only-if *at least one* of the participants in the actual execution would detect the same error. This allows the programmer to reason about failure in the actual distributed execution via the much simpler failure semantics of the single threaded interpretation. (A static type system could be developed to prevent the possibility of failure altogether, but making it suitably expressive would be very challenging, and so remains future work.)

### 3.2 Advanced Coordination

Symphony supports advanced coordination patterns using *first-class shares* and *first-class party sets*, combined with standard functional programming constructs.

**First-class shares** While prior languages support shares as data, they restrict their creation and operations on them, i.e., they are not truly *first class*. Symphony is more flexible, allowing shares to be *delegated* and *reshared*. To support delegation, we can write `share`$[\phi, \tau : \{p\} \rightarrow Q]\ v$ for each IO party $p$ to share its input with compute parties $Q$, who will then compute and share the final result with each $p$. Unlike prior systems, there is no need for $p$ to be a member of $Q$. To support resharing, we give *shares* as inputs $v$ to `share`$[\phi, \tau : P \rightarrow Q]\ v$, rather than plaintext values.

**First-class party sets** Delegation and resharing are made more powerful through the use of *first-class party sets*—sets like `{A,B}` are run-time values, not static annotations. The following Symphony program delegates to parties E and F to compute whether party A is the richest among parties A–D.

```
1  principal A B C D E F
2
3  -- read input at p, delegate to all in Q
4  def readShare Q p = par ({ p } \/ Q)
5    let i = par { p } read int from "input.txt" in -- file local to each p
6    share [gmw, int : { p } -> Q] i
7
8  -- delegate shares from each p in P to all in Q
9  def delegateShares P Q = map (readShare Q) (psetToList P)
10
11 def main () = par {A,B,C,D,E,F}
12   let sharesList = delegateShares {A,B,C,D} {E,F} in
13   let a = head sharesList in
14   let res = par {E,F} fold_list true (fun s res -> res && a >= s)
     ↪ sharesList in
15   reveal [gmw, bool : {E,F} -> {A,B,C,D}] res
```

The `delegateShares` function takes two party sets as inputs: `P` is the set of IO parties, and `Q` is the set of compute parties. The function returns a list of shares, one for each of the parties in `P`, located at parties `Q`. It computes this result via calling `map` over the parties `P`, using `readShare Q` to read an input and share it to `Q`.

Line 12 computes shares for parties `{A,B,C,D}` and delegates them to `{E,F}`. Line 14 then directs `{E,F}` to fold over these shares, checking whether the A's input is consistently the largest. Line 15 reveals the result back to the IO parties.

Not shown here, Symphony allows a party set to be deconstructed via `case`, which is useful for dynamically coordinating among parties, e.g., to elect a committee to carry out a delegated MPC. The located-data invariant allows the choice of parties to be reliably coordinated at run-time.

**LWZ** Resharing, delegation, and first-class party sets are crucial to implementing the efficient *secure shuffle* computation developed by Laur, Willemson, and Zhang [23]. Here is a piece of the Symphony implementation of it (the full code is in Listing 4, Appendix A.2).

```
1  def shuffleWith Q S sharesQ = par (Q \/ S)
2    let sharesS = share [gmw, array int : Q -> S] sharesQ in
3    share [gmw, array int : S -> Q] (shuffle S sharesS)
```

`Q` and `S` are both party sets and `sharesQ` is an array of `int` shares located at `Q`. Line 2 reshares `sharesQ` to parties `S`, and line 3 call `shuffle` to mix up those shares at `S` before resharing the result back to `Q`. Language support for resharing is unique to Symphony and critical for implementing LWZ. To the best of our knowledge, no other MPC language can express this protocol, due to its coordination challenges.

### 3.3 Comparison to Related Work

Symphony is (partly) compared against related MPC frameworks in Table 1. It surpasses them all in terms of expressiveness. Like many frameworks, it supports more than 2 parties, reactive MPC, and synchronized randomness. Like Wysteria, it supports safe coordination via `par` blocks, but generalizes it: programs can *pattern-match* on party sets (not just construct them), and encrypted values can be computed on directly (no need for `sec` blocks), delegated, and reshared. The correspondence theorems we prove between Symphony's distributed and single-threaded semantics (Section 6) are similar to the metatheory results of Wysteria, however the invariants on located data are unique to Symphony and the theorems are adapted to Symphony's dynamic type system. In terms of implementation (Section 7), Symphony supports both garbled circuits (for 2 parties) and secret shares (for 3 or more). All these features together enable good performance.

One related work shown in Table 1 very different from the rest is Viaduct [1], which compiles a Java-like language to secure distributed programs. Viaduct's programming model is higher-level than Symphony; computation/communication patterns are synthesized based on security policies specified as information flow labels [31], rather than specified directly. It also supports cryptographic schemes beyond MPC (e.g., commitments, zero-knowledge proofs). Symphony supports richer coordination patterns than Viaduct (e.g., LWZ), due to its first-class principal sets, bundles, and par blocks. Viaduct has no result corresponding to Symphony's soundness guarantee (Section 6), but has specific means to specify security policies, which may involve declassification and endorsement. Symphony essentially takes an "ideal world" approach, relying on the programmer to ensure a program does not release too much.

Symphony's semantics of "generalized SIMD" bears resemblance to that of *choreography* languages [8, 9, 10, 25]. Choreographic programs are conceptually sequential, ensuring that `send` and `receive` operations are always matched up by combining them into a single expression. Pirouette [18] is a typed choreographic functional programming language which proves that the distributed deployment of well-typed programs is deadlock free by design. Pirouette is able to prove strong metatheoretic properties relating choreographies to their distributed deployment due to its static typing and static party annotations.

## 4 $\lambda$-Symphony: Syntax and Semantics

$\lambda$-Symphony is a minimal core language which captures the essential features of Symphony. This section presents its syntax and *single-threaded* (ST) semantics.

### 4.1 Syntax

The syntax of $\lambda$-Symphony is shown in Fig. 1. To simplify the formal semantics, the syntax adheres to a kind of *administrative normal form* (ANF), meaning that most expression forms operate directly on variables $x$, rather than subexpressions $e$, as is the case in the actual implementation. We isolate *atomic expressions* $a$ as a sub-category of full expressions $e$; the former evaluate to a final result in one "small" step.

Most of the syntactic forms are standard. Binary operations apply either to integers or shares (e.g., $+$, $\times$) or to party sets (e.g., $\cup$). Conditionals $x \mathbin{?} x \diamond x$ correspond to multiplexor expressions (written `mux if` in Symphony). Pairs are accessed via projection (e.g., $\pi_1\ \langle 1, 2\rangle$ evaluates to 1), while sums (aka *variants* or *tagged unions*) are accessed via pattern matching (e.g., $\texttt{case}\ \iota_1\ 0\ \{y.\ e_1\}\ \{y.\ e_2\}$ evaluates to $e_1$ wherein $y$ is substituted with 0). Party sets are also accessed via `case` and processed like lists—the first branch handles the $\varnothing$ case, while the second binds two variables, one for a selected party and the other for the rest of the set. Recursive functions are written $\lambda_z x.\ e$; the function body $e$ may refer to itself via variable $z$. $\lambda$-SYMPHONY also has mutable references, and primitives for I/O. $\lambda$-SYMPHONY does not model lists or bundles because they are easily encoded; we explain how in the Appendix. The MPC-related constructs `par`, `share`, and `reveal` match their SYMPHONY counterparts; the latter two elide the output type and protocol annotation (which are useful for an implementation but unnecessary for formal modeling). SYMPHONY's implementation generalizes other aspects of $\lambda$-SYMPHONY, too, as discussed in Section 7; e.g., it permits sharing values of any type, and doing case analysis on encrypted sums.

| | | | |
|---|---|---|---|
| $i \in$ | $\mathbb{Z}$ | | *integers* |
| $A, B, C \in$ | party | | *parties* |
| $m, p, q \in$ | pset $\triangleq$ | $\wp(\text{party})$ | *sets of parties* |
| $x, y, z \in$ | var | | *variables* |
| $\odot \in$ | bop | | *binary ops (+, ×, ∪, ...)* |
| $a \in$ | atom ::= | $x$ | *variable reference* |
| | \| | $i$ | *integer literal* |
| | \| | $p$ | *party set literal* |
| | \| | $x \odot x$ | *binary operation* |
| | \| | $x\ ?\ x \diamond x$ | *conditional* |
| | \| | $\iota_i\ x$ | *sum injection* |
| | \| | $\langle x, x \rangle$ | *pair creation* |
| | \| | $\pi_i\ x$ | *pair projection* |
| | \| | $\lambda_z x.\ e$ | *(rec.) function def* |
| | \| | $\texttt{ref}\ x$ | *reference creation* |
| | \| | $!x$ | *dereference* |
| | \| | $x := x$ | *reference assignment* |
| | \| | $\texttt{read}$ | *read int input* |
| | \| | $\texttt{write}\ x$ | *write output* |
| | \| | $\texttt{share}[x \rightarrow x]\ x$ | *share encrypted val.* |
| | \| | $\texttt{reveal}[x \rightarrow x]\ x$ | *reveal encrypted val.* |
| $e \in$ | expr ::= | $a$ | *atomic expression* |
| | \| | $\texttt{case}\ x\ \{\bar{x}.e\}\{\bar{x}.e\}$ | *elim for sums, psets* |
| | \| | $x\ x$ | *function call* |
| | \| | $\texttt{par}\ x\ e$ | *parallel execution* |
| | \| | $\texttt{let}\ x = e\ \texttt{in}\ e$ | *let binding* |

**Figure 1** $\lambda$-SYMPHONY formal syntax.

## 4.2 Overview

The ST semantics for $\lambda$-SYMPHONY models all participating parties as if they were executing in lockstep. We prove that the ST semantics faithfully models the *distributed* (DS) semantics presented in Section 5, according to which parties may act inde-

| | | | |
|---|---|---|---|
| $\ell \in \text{loc}$ | | | *memory locations* |
| $\psi \in \text{prot}$ | $::=$ | $\cdot$ | *cleartext* |
| | $\mid$ | $\texttt{enc}\#m$ | *encrypted* |
| $\gamma \in \text{env}$ | $\triangleq$ | $\text{var} \rightharpoonup \text{value}$ | *value environment* |
| $\delta \in \text{store}$ | $\triangleq$ | $\text{loc} \rightharpoonup \text{value}$ | *value store* |
| $u \in \text{loc-value}$ | $::=$ | $i^{\psi}$ | *integer/share value* |
| | $\mid$ | $p$ | *party set value* |
| | $\mid$ | $\iota_i\ v$ | *tagged union injection* |
| | $\mid$ | $\langle v, v\rangle$ | *pairs* |
| | $\mid$ | $\langle \lambda_z x.\ e, \gamma\rangle$ | *closures* |
| | $\mid$ | $\ell^{\#m}$ | *reference* |
| $v \in \text{value}$ | $::=$ | $u@m$ | *located value* |
| | $\mid$ | $\bigstar$ | *opaque value* |

$$\boxed{_\surd_m \in \text{loc-value} \to \text{loc-value}}$$

$$\bigstar\surd_m \triangleq \bigstar \qquad\qquad (u@p)\surd_m \triangleq \begin{cases} u\surd_m@(p \cap m) & \text{if } p \cap m \neq \varnothing \\ \bigstar & \text{if } p \cap m = \varnothing \end{cases}$$

**Figure 2** $\lambda$-Symphony definitions and metafunctions used in formal semantics (partial).

pendently. Thus, the ST semantics can serve as the basis of $\lambda$-Symphony formal reasoning, e.g., about correctness and security.

The main judgment $\varsigma \longrightarrow \varsigma$ is a reduction relation between *configurations* $\varsigma$. A configuration is a 5-tuple comprising the current mode $m$, environment $\gamma$, store $\delta$, stack $\kappa$, and expression $e$. The mode is the set of parties computing $e$ in parallel; we say the parties $A \in m$ are *present* for a computation. Per Figure 2, environments are partial maps from variables to values, and stores are partial maps from memory locations to values; we discuss stacks shortly. The main judgment employs $\gamma \vdash_m \delta, a \hookrightarrow \delta, v$, which defines the reduction of atomic expressions $a$ to values $v$. Selected rules for both judgments are given in Figure 3; the full set of rules appears in the Appendix.

### 4.3 Values

*Values* $v$ have one of two forms: $u@m$ indicates that the *located value* $u$ is only accessible to $A \in m$, e.g., because it was the result of evaluating $e$ in mode $m$; whereas $\bigstar$ is the *opaque value* which is both unknown and inaccessible. Located values are defined in Figure 2, including for numbers $i^{\psi}$, sets of parties $p$, sums $\iota_i\ v$, pairs $\langle v, v\rangle$, recursive functions $\langle \lambda_z x.\ e, \gamma\rangle$ which include a closure environment $\gamma$, and memory locations (i.e., pointers) $\ell^{\#p}$. These are standard except for annotations $\psi$ and $\#p$.

The annotation $\#p$ to indicate the parties $p$ that are *co-creators* of the referenced memory, whereas $\psi$ indicates the *protocol* of the annotated integer: $\cdot$ represents a cleartext value (we write just $i$ when the annotation $\psi$ is $\cdot$), whereas $\texttt{enc}\#p$ represents an encrypted value shared among parties $B \in p$ (a "share"). Thus, a value $1^{\texttt{enc}\#q}@q$

$$\boxed{\gamma \vdash_m \delta, a \hookrightarrow \delta, v}$$

ST-VAR
$$\frac{}{\gamma \vdash_m \delta, x \hookrightarrow \delta, \gamma(x)\surd_m}$$

ST-LIT
$$\frac{}{\begin{array}{l}\gamma \vdash_m \delta, i \hookrightarrow \delta, i@m \\ \gamma \vdash_m \delta, p \hookrightarrow \delta, p@m\end{array}}$$

ST-INT-BINOP
$$\frac{i_1^{\psi}@m = \gamma(x_1)\surd_m \qquad i_2^{\psi}@m = \gamma(x_2)\surd_m \qquad \vdash_m \psi}{\gamma \vdash_m \delta, x_1 \odot x_2 \hookrightarrow \delta, [\![\odot]\!](i_1, i_2)^{\psi}@m}$$

ST-REF
$$\frac{v = \gamma(x)\surd_m}{\gamma \vdash_m \delta, \texttt{ref}\ x \hookrightarrow \{\ell \mapsto v\} \uplus \delta, \ell^{\#m}@m}$$

ST-DEREF
$$\frac{\ell^{\#q}@m = \gamma(x)\surd_m}{\gamma \vdash_m \delta, !x \hookrightarrow \delta, \delta(\ell)\surd_m}$$

ST-ASSIGN
$$\frac{\ell^{\#m}@m = \gamma(x_1)\surd_m \qquad v = \gamma(x_2)\surd_m}{\gamma \vdash_m \delta, x_1 := x_2 \hookrightarrow \delta[\ell \mapsto v], v}$$

ST-READ
$$\frac{|m| = 1}{\gamma \vdash_m \delta, \texttt{read} \hookrightarrow \delta, i@m}$$

ST-WRITE
$$\frac{i@m = \gamma(x)\surd_m \qquad |m| = 1}{\gamma \vdash_m \delta, \texttt{write}\ x \hookrightarrow \delta, 0@m}$$

ST-SHARE
$$\frac{\begin{array}{ll} p@m = \gamma(x_1)\surd_m & \vdash_p \psi \\ q@m = \gamma(x_2)\surd_m & q \neq \varnothing \\ i^{\psi}@p = \gamma(x_3)\surd_p & m = p \cup q \end{array}}{\gamma \vdash_m \delta, \texttt{share}[x_1 \rightarrow x_2]\ x_3 \hookrightarrow \delta, i^{\texttt{enc}\#q}@q}$$

ST-REVEAL
$$\frac{\begin{array}{ll} p@m = \gamma(x_1)\surd_m & \\ q@m = \gamma(x_2)\surd_m & q \neq \varnothing \\ i^{\texttt{enc}\#p}@p = \gamma(x_3)\surd_p & m = p \cup q \end{array}}{\gamma \vdash_m \delta, \texttt{reveal}[x_1 \rightarrow x_2]\ x_3 \hookrightarrow \delta, i@q}$$

$$\boxed{\varsigma \longrightarrow \varsigma}$$

ST-PAR
$$\frac{p@m = \gamma(x)\surd_m \qquad m \cap p \neq \varnothing}{m, \gamma, \delta, \kappa, \texttt{par}\ x\ e \longrightarrow m \cap p, \gamma, \delta, \kappa, e}$$

ST-PAREMPTY
$$\frac{p@m = \gamma(x)\surd_m \qquad m \cap p = \varnothing \qquad \gamma' = \{x' \mapsto \bigstar\} \uplus \gamma}{m, \gamma, \delta, \kappa, \texttt{par}\ x\ e \longrightarrow m, \gamma', \delta, \kappa, x'}$$

**Figure 3** $\lambda$-SYMPHONY single-threaded semantics, selected rules.

can be read as "an integer 1, encrypted (i.e., secret shared) between parties $q$, and accessible to parties $q$." The first $q$ represents *among whom is this value shared* (determined when the share is created), and the second $q$ represents *who has access to this value* (determined by the enclosing `par` blocks). Location annotations are only used in the ST semantics in order to simulate the presence of multiple parties; they are unused in the distributed semantics and final execution. On the other hand, the $\texttt{enc}\#q$ and $\#p$ annotations are used during distributed execution to detect buggy programs which fail to coordinate properly, e.g., if $A$ alone attempts to do arithmetic on a share owned by both $A$ and $B$, or if only $A$ attempts to write to a reference it co-created with $B$.

### 4.4 Operational Rules

Now we consider some of the operational rules.

**Literals, Variables, Binding, Computation** Rule ST-VAR retrieves a variable's value from the environment and *(re)locates it* to the current mode $m$ via $\gamma(x)\not\downarrow_m$. The function $_\not\downarrow_m$ is given in Figure 2. For values $u@p$, $_\not\downarrow_m$ relocates them to $p \cap m$, unless the intersection is empty in which case the value is inaccessible, so it becomes $\bigstar$. Relocating is a deep operation; $u@p\not\downarrow_m$ also relocates the contents $u$ to $u\not\downarrow_m$, which recurses over the sub-terms of $u$. Relocation ensures that the retrieved value is *compatible with* $m$. A value $v$ is compatible with a set of parties $m$ when it is accessible to some set of parties $p \subseteq m$. Compatibility with the current mode is a general invariant of all of the rules. Rule ST-LIT types integer and principal literals, annotating them with (compatible) location $m$. The full definition of relocation appears in the Appendix.

Variables bound with `let` are managed using a stack $\kappa$, which is either the empty stack $\top$ or a list of frames $\langle \texttt{let}\ x = \square\ \texttt{in}\ e \mid m, \gamma \rangle :: \kappa$. To evaluate $\texttt{let}\ x = e_1\ \texttt{in}\ e_2$, we push frame $\langle \texttt{let}\ x = \square\ \texttt{in}\ e_2 \mid m, \gamma \rangle$ and set the active expression to $e_1$ (Rule ST-LET-PUSH, not shown). When an expression evaluates to a value $v$, the topmost frame $\langle \texttt{let}\ x = \square\ \texttt{in}\ e_2 \mid m, \gamma \rangle$ is popped and evaluation proceeds on $e_2$ using saved mode $m$ and environment $\gamma$ updated to map $x$ to $v$ (Rule ST-LET-POP, not shown).

Rule ST-INT-BINOP handles arithmetic over integers. This rule illustrates another invariant that all elimination rules share. To compute on a value while running in mode $m$ requires that the value be accessible to all parties in $m$. We see this in premises like $i_1^{\psi}@m = \gamma(x_1)\not\downarrow_m$, which locate the operated-on variable to current mode $m$ and then ensure that the value's location is also $m$, i.e., all parties have access to the computed-on value. Doing so ensures that these parties, when running in a distributed setting with their own store, environment, etc. will agree on the result. For this rule in particular, we also ensure that both integers have the same protocol $\psi$, and that this protocol is compatible with mode $m$, written $\vdash_m \psi$. Compatibility holds when $\psi$ is cleartext, and when it is $\texttt{enc}\#m$, i.e., $i$ is a share amongst all parties currently present. The distributed semantics also uses compatibility checks to ensure parties are in sync.

**Par mode** Operationally, $\texttt{par}\ x\ e$ evaluates $e$ in mode $m \cap p$ where $p@m = \gamma(x)\not\downarrow_m$; i.e., only those parties in $p$ which are *also* present in $m$ will run $e$. When $m \cap p$ is non-empty, rule ST-PAR directs $e$ to evaluate in the refined mode. If $m \cap p$ is empty, then per rule ST-PAREMPTY, $e$ is skipped and $\bigstar$ is returned.[2] Note that because the stack tracks each frame's mode, when the current expression completes the old mode will be restored when a stack frame is popped.

Here is an example of how par mode and variable access interact.

```
par {A,B} let x = par {A} 1 in
          let y = par {B} x in
          let z = par {C} 2 in x
```

The outer $\texttt{par}\ \{A, B\}$ evaluates its body in mode $\{A, B\}$, per rule ST-PAR. Next, according to rules ST-LETPUSH, ST-PAR, and ST-INT we evaluate 1 in mode $m = \{A, B\} \cap \{A\} = \{A\}$; we bind the result $1@\{A\}$ to $x$ in $\gamma$ per rule ST-LETPOP. Next, according to rules ST-

[2] Since $\bigstar$ is not an expression—it is a value—we return a fresh variable and the environment with that variable mapped to $\bigstar$.

LETPUSH, ST-PAR and ST-VAR we evaluate $x$ in mode $m = \{B\}$. Per rule ST-VAR, we retrieve value $1@\{A\}$ for $x$, and then $\surd_{\{B\}}(1@\{A\})$ yields ★ as the result, which is bound to $y$ in $\gamma$ per rule ST-LETPOP. This result makes sense: Party $B$ reads variable $x$ whose contents are only accessible to $A$, so all it can do is return the opaque value. Finally, par $\{C\}$ 2 evaluates to ★ according to rule T-PAREMPTY, since $m = \{A,B\} \cap \{C\} = \varnothing$. This ★ result is bound to $z$ per rule ST-LETPOP, and the final result $x$, evaluated in mode $m = \{A,B\}$ is $\surd_{\{A,B\}}(1@\{A\}) = 1@\{A\}$ per rule ST-VAR.

**References** Rule ST-REF creates a fresh reference and returns a located pointer annotated with the parties that created it. Rule ST-DEREF takes a reference located in the current mode $m$ and returns the pointed-to contents made compatible with $m$. Rule ST-ASSIGN updates the store with the new value and returns it, but only works for $\ell^{\#p}$ references where $p = m$, the current mode. Why? Consider the following example.

```
par {A,B} let x = ref 0 in
          let _ = (par{A} x := 1) in
          let y = !x in ...
```

The variable $x$ initially contains a reference $\ell^{\#\{A,B\}}$ because it was created in a context with mode $m = \{A,B\}$. Then $x$ is assigned to by $A$ in the par expression on the subsequent line. By rule ST-ASSIGN, the creators of the reference $\#\{A,B\}$ must match mode $m$ to proceed, but since $m$ is $\{A\}$ the program is stuck. This is desirable because to proceed would cause $A$'s and $B$'s views of the computation to get out of sync. When we run this program at each of $A$ and $B$ separately, as part of the distributed semantics, on $A$ we would do the assignment but on $B$ it would be skipped. As such, on $A$ the value of $y$ would be 1 but on $B$ it would be 0. If the continuation of the program in ... were to branch on $y$ and then in one branch do some MPC constructs but not in the other, then the two parties would become even further out of sync.

**I/O** Rules ST-READ and ST-WRITE handle I/O. They require the mode to be a singleton party, and locate the resulting value at that singleton party. This is important for ensuring that all parties agree on the contents of shared variables. In the semantics, `read` nondeterministically returns any integer, which over-approximates the behavior of reading a particular integer as input from the host environment.

**Multiparty computation** Party $A$ creates encrypted values (i.e., shares) among parties $q$ using syntax $\texttt{share}[x_1 \rightarrow x_2]\ x_3$ handled by rule ST-SHARE. Variable $x_1$ is the set of input parties $p$; variable $x_2$ is the (nonempty) set of parties who will hold shares $q \neq \varnothing$; and $x_3$ is the value to be shared, known to the input parties $p$; if $x_3$ is already an encrypted share (among $p$) then this operation will *reshare* it among $q$. The input parties $p$ and share parties $q$ must all be present in the mode $m$, and no other parties may be present (so $m = p \cup q$). Importantly, there is *no requirement* that $p \subseteq q$—this means that parties $p$ may *delegate* the secure computation to parties $q$ and not participate in it themselves. Both resharing and delegation are key strengths of SYMPHONY not present in existing MPC languages. The `share` expressions in LWZ (Listing 4, Appendix A.2) (critically) leverage both of these features.

$$\dot v \in \text{lval} ::= i^{\psi} \mid p \mid \ell^{\#m} \mid \iota_i\ \dot v \mid \langle \dot v, \dot v \rangle \mid \langle \lambda_z x.\ e, \dot\gamma \rangle \mid \bigstar \qquad \dot\varsigma \in \text{lconfig} ::= m, \dot\gamma, \dot\delta, \dot\kappa, e \qquad C \in \text{dconfig} \triangleq \text{party} \rightharpoonup \text{lconfig}$$

$$\boxed{\dot\gamma \vdash_m \dot\delta, a \hookrightarrow \dot\delta, \dot v}$$

$$\textsc{DS-Var}\ \frac{}{\dot\gamma \vdash_m \dot\delta, x \hookrightarrow \dot\delta, \dot\gamma(x)} \qquad \textsc{DS-Int-Binop}\ \frac{i_1^{\psi} = \dot\gamma(x_1) \quad i_2^{\psi} = \dot\gamma(x_2) \quad \vdash_m \psi}{\dot\gamma \vdash_m \dot\delta, x_1 \odot x_2 \hookrightarrow \dot\delta, [\![\odot]\!](i_1, i_2)^{\psi}}$$

$$\boxed{\dot\varsigma \longrightarrow_A \dot\varsigma}$$

$$\textsc{DS-Par}\ \frac{p = \dot\gamma(x) \quad A \in p}{m, \dot\gamma, \dot\delta, \dot\kappa, \texttt{par}\ x\ e \longrightarrow_A m \cap p, \dot\gamma, \dot\delta, \dot\kappa, e} \qquad \textsc{DS-ParEmpty}\ \frac{p = \dot\gamma(x) \quad A \notin p \quad \dot\gamma' = \{x' \mapsto \bigstar\} \uplus \dot\gamma}{m, \dot\gamma, \dot\delta, \dot\kappa, \texttt{par}\ x\ e \longrightarrow_A m, \dot\gamma', \dot\delta, \dot\kappa, x'}$$

$$\boxed{C \rightsquigarrow C}$$

$$\textsc{DS-Step}\ \frac{\dot\varsigma \longrightarrow_A \dot\varsigma'}{\{A \mapsto \dot\varsigma\} \uplus C \rightsquigarrow \{A \mapsto \dot\varsigma'\} \uplus C}$$

$$\textsc{DS-Share}\ \frac{\begin{array}{lll} \texttt{share}[x_1 \to x_2]\ x_3 = C(m).e & \vdash_p \psi & C' = \{A \mapsto (m, \{x \mapsto \dot v\} \uplus \dot\gamma, \dot\delta, \dot\kappa, x) \\ p = C(m).\dot\gamma(x_1) & m = C(m).m & \quad \mid C(A) = (m, \dot\gamma, \dot\delta, \dot\kappa, e), \\ q = C(m).\dot\gamma(x_2) & q \neq \varnothing & \quad A \in q \implies \dot v = i^{\mathsf{enc}\#q}, \\ i^{\psi} = C(p).\dot\gamma(x_3) & m = p \cup q & \quad A \in p \wedge A \notin q \implies \dot v = \bigstar\} \end{array}}{C_0 \uplus C \rightsquigarrow C_0 \uplus C'}$$

$$\textsc{DS-Reveal}\ \frac{\begin{array}{lll} \texttt{reveal}[x_1 \to x_2]\ x_3 = C(m).e & & C' = \{A \mapsto (m, \{x \mapsto \dot v\} \uplus \dot\gamma, \dot\delta, \dot\kappa, x) \\ p = C(m).\dot\gamma(x_1) & m = C(m).m & \quad \mid C(A) = (m, \dot\gamma, \dot\delta, \dot\kappa, e), \\ q = C(m).\dot\gamma(x_2) & q \neq \varnothing & \quad A \in q \implies \dot v = i, \\ i^{\mathsf{enc}\#p} = C(p).\dot\gamma(x_3) & m = p \cup q & \quad A \in p \wedge A \notin q \implies \dot v = \bigstar\} \end{array}}{C_0 \uplus C \rightsquigarrow C_0 \uplus C'}$$

**Figure 4** λ-Symphony distributed semantics, selected rules. Note that judgment $\dot\gamma \vdash_m \dot\delta, a \hookrightarrow \dot\delta, \dot v$ is referred to by the elided rule DS-LetPop of the $\dot\varsigma \longrightarrow_A \dot\varsigma$ judgment, analogously to the single threaded semantics in Fig. 3.

Per rule ST-Reveal, $\texttt{reveal}[x_1 \to x_2]\ x_3$ reveals a shared encrypted value among parties $p$ to a cleartext result at parties $q \neq \varnothing$, where $x_1$ evaluates to $p$, $x_2$ evaluates to $q$, and $x_3$ evaluates to the encrypted value. All parties $p$ among which $x_3$ is shared must be present, as well as the recipients of the value $q$, and no other parties.

## 5 Distributed Semantics

This section presents λ-Symphony's *distributed* (DS) semantics, modeling the communication and coordination needed for MPC. The next section proves the correspondence of the ST semantics w.r.t. the DS semantics.

### 5.1 Configurations

A *distributed configuration* $C$ collects the execution states of the individual parties in an MPC. As shown at the top of Figure 4, it consists of a finite map from parties to *local configurations* $\dot\varsigma$, which are 5-tuples consisting of (1) a mode $m$, (2) a local environment $\dot{\gamma}$, (3) a local store $\dot{\delta}$, (4) a local stack $\dot{\kappa}$, and (5) an expression $e$. Local environments, stores, and stacks are the same as their ST counterparts except that instead of containing values $v$, they contain *local values* $\dot{v}$, which lack location annotations $@m$.

For a set of parties $m$ wishing to jointly execute program $e$, the initial configuration $C_0$ will map each party $A \in m$ to a local configuration $(m, \varnothing, \varnothing, \top, e)$, where $\varnothing$ is the empty function (used for the empty environment and store), $\top$ is the empty stack, and $e$ is the source program. Notice that each party tracks the *global* mode $m$ in its local configuration.

### 5.2 Operational Semantics

The DS semantics $C \rightsquigarrow C'$ uses auxiliary judgments $\dot{\gamma} \vdash_m \dot{\delta}, a \hookrightarrow \dot{\delta}, \dot{v}$ and $\dot{\varsigma} \longrightarrow_A \dot{\varsigma}$; these are defined in part in Figure 4. The main rule DS-STEP is used to execute a single party, independently of the rest. The rule selects some party $A$'s local configuration $\dot{\varsigma}$, steps it to $\dot{\varsigma}'$, and then incorporates that back into the distributed configuration. This rule can be used whenever $A$'s active expression $e$ is anything other than `share` or `reveal`, which require synchronizing between multiple parties. Those two cases use the rules DS-SHARE and DS-REVEAL, respectively, discussed below.

**Non-synchronizing expressions** The rules for relation $\dot{\gamma} \vdash_m \dot{\delta}, a \hookrightarrow \dot{\delta}, \dot{v}$ are essentially the same as those for the ST semantics, except that they operate on non-located data. The figure shows two examples. Rule ST-VAR's conclusion locates the result at $m$ via $\gamma(x)\not{}_m$, but rule DS-VAR's conclusion simply returns $\dot{\gamma}(x)$. Similarly, rule ST-INTBINOP's premise requires $i_1^{\psi}@m = \gamma(x_1)\not{}_m$ while rule DS-INTBINOP's premise simply requires $i_1^{\psi} = \dot{\gamma}(x_1)$. For elimination forms, a location mismatch in a ST rule would translate to failed attempt to eliminate $\bigstar$ in the DS rule. For example, if rule ST-INTBINOP would have failed because $i_1^{\psi}$ was located not at $m$ but at $p \subset m$ instead, then rule DS-INTBINOP would fail for parties $A \in (m - p)$ since for these $\dot{\gamma}(x_1) = \bigstar$, which cannot be added to another share. The check $\vdash_m \psi$ is present in both rules to prevent attempts to add incompatible shares. Likewise, rules DS-DEREF and DS-ASSIGN (not shown) retain the check from the ST versions that the reference owners are compatible with $m$.

Judgment $\dot{\varsigma} \longrightarrow_A \dot{\varsigma}$ corresponds to ST judgment $\varsigma \longrightarrow \varsigma$, where annotation $A$ indicates the executing local party. The rules for both judgments are essentially the same except for those handling `par[`$x$`]` $e$. Rule DS-PAR evaluates to the expression $e$ so long as $A \in p$, where $p = \dot{\gamma}(x)$, updating the global mode to $m \cap p$, just as the ST semantics does. Rule DS-PAREMPTY handles the case when $A \notin p$, thus skipping $e$ and leaving global mode $m$ as it is, evaluating to result $x'$, which is a fresh variable bound to $\bigstar$ in $\dot{\gamma}'$.

**Synchronizing expressions** Rules DS-SHARE and DS-REVEAL are used to evaluate expressions `share` and `reveal`, respectively. These expressions require synchronizing between multiple parties, transferring data from one party to the other(s), so the rules manipulate multiple local configurations. But they are quite similar to their ST counterparts.

$$\boxed{_\lightning_A \in \text{loc-value} \to \text{lvalue} \; ; \; \text{value} \to \text{lvalueenv} \to \text{lenv} \; ; \; \text{store} \to \text{lstore} \; ; \; \text{stack} \to \text{lstack}}$$

$$\begin{array}{c} i^{\psi}\lightning_A \triangleq i^{\psi} \\ p\lightning_A \triangleq p \\ (\iota_i\ v)\lightning_A \triangleq \iota_i\ v\lightning_A \end{array} \qquad \begin{array}{c} \langle v_2, v_2\rangle\lightning_A \triangleq \langle v_1\lightning_A, v_2\lightning_B\rangle \\ \langle \lambda_z x.\ e, \gamma\rangle\lightning_A \triangleq \langle \lambda_z x.\ e, \gamma\lightning_A\rangle \\ \ell^{\#m}\lightning_A \triangleq \ell^{\#m} \end{array} \qquad \begin{array}{c} u@p\lightning_A \triangleq \begin{cases} u\lightning_A & \mathit{if}\ \ A \in p \\ \bigstar & \mathit{if}\ \ A \notin p \end{cases} \\ \bigstar\lightning_A \triangleq \bigstar \end{array} \qquad \begin{array}{c} \gamma\lightning_A(x) \triangleq \gamma(x)\lightning_A \\ \delta\lightning_A(\ell) \triangleq \delta(\ell)\lightning_A \end{array}$$

$$\top\lightning_A \triangleq \top \qquad (\langle \texttt{let}\ x = \square\ \texttt{in}\ e \mid m, \gamma\rangle :: \kappa)\lightning_A \triangleq \langle \texttt{let}\ x = \square\ \texttt{in}\ e \mid m, \gamma\lightning_A\rangle :: \kappa\lightning_A$$

$$(m, \gamma, \delta, \kappa, e)\lightning \triangleq \{A \mapsto m, \gamma\lightning_A, \delta\lightning_A, \kappa\lightning_A, e \mid A \in m\} \qquad \boxed{_\lightning \in \text{config} \to \text{dconfig}}$$

**Figure 5** Slicing metafunction; relates ST and DS semantics.

In the rules we write $C(m)$ to refer to the set of configurations mapped to by principals $A \in m$. When we write $C(m).e = e'$, we are saying that the expression component ($e$) of each configuration in the set $C(m)$ is equal to expression $e'$. For DS-Share, $e'$ is $\texttt{share}[x_1 \to x_2]\ x_3$ and for DS-Reveal it is $\texttt{reveal}[x_1 \to x_2]\ x_3$. We similarly insist that each party's configuration agrees on the valuation of $x_1$ to $p$ and $x_2$ to $q$, which together comprise the agreed-upon mode $m$. For DS-Share, the valuation of $x_3$ must be an integer with a protocol $\psi$ compatible with $p$: $\vdash_p \psi$; for DS-Reveal, the valuation of $x_3$ for all sharing parties $p$ must be an encrypted integer shared amongst those parties. These conditions are sufficient to guarantee that the *share* and *reveal* operations of the actual MPC backend complete successfully.[3] The updated configuration $C'$ matches the original configuration $C$ but updates the local configuration for each party $A \in m$ to have expression component $x$, where $x$ is a fresh variable added to the store $\dot{\gamma}$ to map to the communicated (cleartext or encrypted) integer; those sharing parties $A \in p$ such that $A \notin q$ evaluate to $\bigstar$ instead.

## 6 Single-Threaded Soundness

This section presents our main meta-theoretical results around *single-threaded soundness*, which is the sense in which we can interpret a $\lambda$-Symphony program in terms of its ST semantics, even though in reality it will execute in a distributed fashion. Proofs are provided in Appendix B.2.

We relate a single-threaded configuration $\varsigma$ to a distributed one by *slicing* it, written $\varsigma\lightning$, which is defined in Figure 5. Each party $A$ in the mode $m$ of $\varsigma$ is mapped to its local DS configuration consisting of $m$, expression $e$, and the sliced versions of the environment $\gamma$, store $\delta$, and stack $\kappa$ of $\varsigma$ that are specific to $A$. Slicing captures the simple idea that for a value $u@p$, if $A \in p$ then $A$ can access $u$, but if $A \notin p$ then it cannot; $_\lightning_A$ works much like $_\checkmark_{\{A\}}$ but strips off all location annotations.

We can prove a full correspondence for programs whose execution trace concludes in normal form that is a *terminal state*. A terminal ST state has an empty stack and has reached a value; a DS state is terminal if all of its local configurations are terminal (see the Appendix. for a formal definition).

[3] Note that in the actual implementation, each party $A \in m$ merely needs to check that its own view of $m$, $p$, and $q$ is consistent per $m = p \cup q$—if not, as shown in the next section, it has landed in a *stuck configuration* and can signal that MPC has failed with a type error.

**Theorem 6.1** (ST/DS Terminal Correspondence). *If $\varsigma \longrightarrow^* \varsigma'$, then statements (1) and (2) imply one another for any $C$:*

*1. $\varsigma'$ is a terminal state and $C = \varsigma' \lightning$* *2. $\varsigma \lightning \rightsquigarrow^* C$ and $C$ is a terminal state*

The proof follows from a *forward simulation* lemma, which establishes that for every single-threaded execution there exists a compatible distributed one, and *confluence*, which establishes that even though the distributed semantics is nondeterministic, its final states are uniquely determined. We additionally prove two lemmas about executions that diverge or get stuck.

**Theorem 6.2** (ST/DS Strong Asymmetric Non-terminal Correspondence). *The following statements are true:*

1. *If $\varsigma \lightning$ reaches a stuck state (under $\rightsquigarrow$) then $\varsigma$ reaches a stuck state (under $\longrightarrow$)*
2. *If $\varsigma$ divergent (under $\longrightarrow$) then $\varsigma \lightning$ divergent (under $\rightsquigarrow$)*

Theorem 6.2 does not rule out the possibility that $\varsigma$ gets stuck while $\varsigma \lightning$ never does. Consider the program `let` $x =$ `par`$[A]$ `<error>` `in` `par`$[B]$ `<infinite loop>`. In the ST semantics this program gets stuck. In the DS semantics, $A$ will only become *locally stuck* while $B$ runs forever.[4] We prove that if an ST configuration gets stuck, then for any reachable DS configuration there exists a reachable locally stuck state:

**Theorem 6.3** (ST/DS Soundness for Stuck States). *If $\varsigma \longrightarrow^* \varsigma'$ and $\varsigma'$ is stuck, then for every $C$ where $\varsigma \lightning \rightsquigarrow^* C$ there exists a $C'$ s.t. $C \rightsquigarrow^* C'$ and $C'$ is locally stuck.*

It follows that if the ST semantics applied to $\varsigma$ detects a runtime error (i.e., gets stuck), then (assuming a non-pathological scheduler) one of the local configurations of $\varsigma \lightning$ will eventually detect a runtime failure (i.e., get locally stuck), too, at which point an implementation could notify the other configurations of the problem.

In sum: the ST and DS semantics correspond for both terminating and non-terminating programs, but with a local notion of "stuckness" applied to DS states.

## 7 Implementation

We implemented a SYMPHONY interpreter in 4K lines of Haskell. The interpreter can run programs in sequential mode for prototyping and debugging, and distributed mode for actual MPC. SYMPHONY adds a number of features beyond $\lambda$-SYMPHONY, including booleans and a conditional expression; nested pattern-matching on pairs, sums, lists, principal sets and bundles; arrays (mutable vectors with O(1) lookup); synchronized randomness; and implicit embedding of constants as shares. SYMPHONY supports semi-honest MPC over boolean circuits, supporting Yao's 2-party protocol with EMP toolkit [36] and the $N$-party GMW protocol with MOTION [5].

[4] Examples like this are also the reason we cannot prove *full bisimulation*.

The Symphony interpreter also generalizes $\lambda$-Symphony by allowing `mux`, `case`, `share` and `reveal` to operate recursively over on pairs, sums and lists. It adds `mux-case` for case analysis on encrypted sums, which are represented with pairs: $\lambda$-Symphony value $\iota_0\ v$ is represented as $\langle \texttt{true}, \langle v, \texttt{default} \rangle \rangle$ and value $\iota_1\ v$ is represented as $\langle \texttt{false}, \langle \texttt{default}, v \rangle \rangle$, with each of the components encrypted. The value `default` is to allow case analysis to proceed on *both* branches of `mux-case`, as a kind of multiplexor. The precise value of `default` is determined when sharing, based on the type annotation $\tau$ on $\texttt{share}[\phi, \tau{:}P \to Q]\ \texttt{default}$.

We have implemented a standard library (about 800 LOC) for Symphony that includes various data structures and coordination patterns, e.g., initializing a bundle from a principal set, and bounded recursion for unrolling an MPC function. Appendix A.1 has more details about the implementation and the standard library's contents.

**Symphony Runtime** EMP and MOTION don't support delegation or resharing, and MOTION does not support reactive MPC. The Symphony runtime, implemented in 2K lines of Rust, acts as a compatibility layer that enhances MPC backends with necessary features. The runtime adds support for delegation and resharing through semi-honest XOR sharing over Symphony parties and support for reactive MPC by caching encrypted values.

Encrypted values in Symphony are represented *abstractly*. The values encrypted with each protocol satisfy the Haskell equivalent of the following Java-ish interface.

```
1  public interface Enc <T> {
2    type Context;
3
4    // Turn the XOR share `share` into an abstract, encrypted `T`
5    void reflect(Context c, BaseValue share);
6
7    // Embed the cleartext constant `clear` as a `T`
8    void constant(Context c, BaseValue clear);
9
10   // Primitive Operations
11   T prim(Context c, Operation op, List<T> shares);
12
13   // Turn the abstract, encrypted `T` value, `enc`, into an XOR share
14   BaseValue reify(Context c, T enc);
15 }
```

To add a new protocol to Symphony, one need only provide a new implementation of this interface. The `type Context` declaration on line 2 is not legal Java, and is meant to be evocative of an associated type in Rust or a type family in Haskell. Each protocol uniquely determines a type as its `Context`. This interface allows the interpreter to ignore differences between MPC based on secret sharing and circuit garbling. For example, the MOTION backend implements MPC lazily by building up a circuit and executing it when a decryption is requested. In contrast, EMP implements MPC eagerly by garbling and executing gates as they are created. Both of these approaches, however, can be made to implement the interface above. Adding the enhancements necessary to do so is the subject of the remainder of this section.

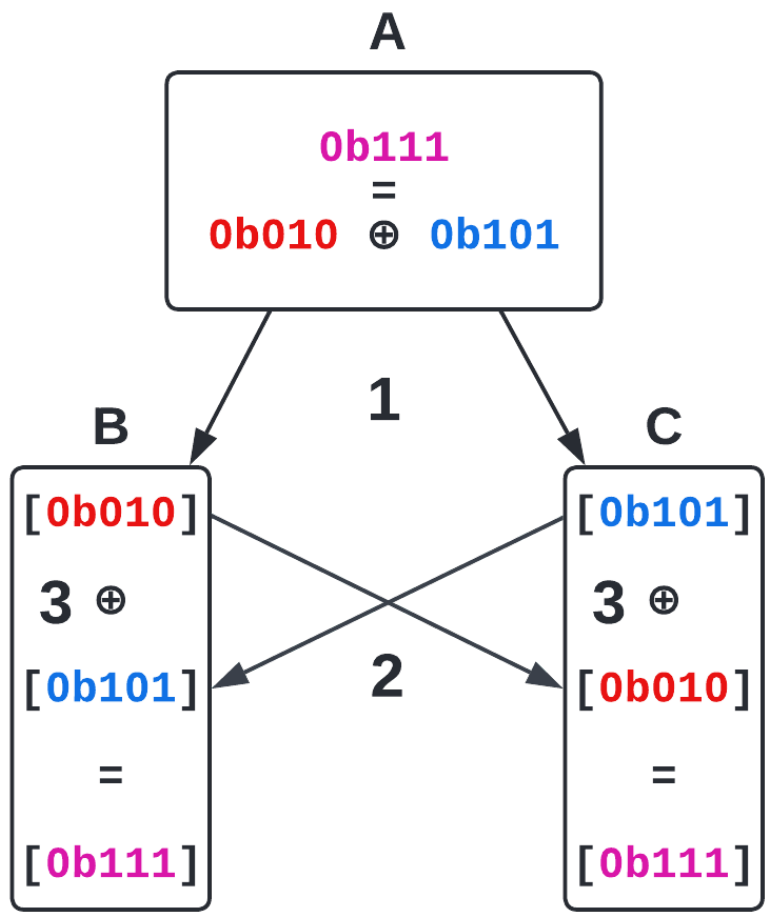

**Figure 6** Runtime support for delegation.

**Delegation and Resharing** The runtime adds support for delegation and resharing through semi-honest XOR sharing over SYMPHONY parties. For example, Figure 6 shows how a party $A$ would delegate their secret bitstring `0b111` to $\{B, C\}$. First, $A$ generates two random XOR shares of her input, in this case `0b010` and `0b101`, and sends them to $B$ and $C$. Second, $B$ and $C$ convert their shares into the native encrypted representation of the backend. We denote a value encrypted in the backend by square brackets, e.g. `[0b101]`. Third, $B$ and $C$ use the MPC backend to compute the XOR of their encryped shares, yielding $A$'s original input natively encrypted among $\{B, C\}$. The resharing operation can be achieved in a similar way, converting natively encryped values to XOR shares and then following the delegation procedure outlined above.

The delegation and resharing procedures are *generic*, treating the underlying MPC backend as a black box and relying only on standard features. However, implementing the procedures generically also fails to take advantage of optimizations that are available when converting between specific protocols. SYMPHONY does not implement every optimal conversion procedure, but there is nothing in our approach preventing us from doing so. Efficient conversion procedures for 2-party and $N$-party protocols are well-documented in the MPC literature [12, 5].

**Reactive MPC** The SYMPHONY runtime also adds support for reactive MPC to the MOTION backend. An MPC context in MOTION, called a `Party`, is a C++ object that manages the global state associated with the MPC. It contains information about the current executing party, the total number of parties, how parties can be contacted (e.g. via TCP sockets), a shared PRG, and a binary circuit composed of gates and wires. When an encrypted value is created or computed, the `Party` object creates a new gate which is added to the circuit. The actual encrypted value is a reference to the output wire of the new gate. When the programmer has finished their MPC, they instruct the `Party` object to execute the underlying circuit. At that point, the circuit is executed using the GMW protocol and XOR shares can be extracted from encrypted values.

Unfortunately, after the `Party` is executed, MOTION does not allow us to reuse the `Party` object for additional MPC: the `Party` object is effectively defunct. The ability to continue performing MPC is precisely what is needed to support reactive MPC, in which values are decrypted and then used to influence additional computation.

The runtime adds support for reactive MPC to MOTION by caching the XOR shares resulting from one execution of a `Party` object, destroying it, and then creating a new one. When we compute on encrypted values produced by the previous `Party`, we provide the corresponding cached XOR shares to the new `Party` as input before proceeding with the computation. While conceptually simple, it required considerable engineering to achieve acceptable performance. For example, we had to modify

**Table 2** A collection of implemented MPC programs (described Appendix A.2 in detail) . # indicates the number of parties. **Features** indicates features required to implement the program: `P` for first-class party sets, `R` for reactive MPC, `$` for synchronized randomness, `D` for delegation, and `S` for resharing. **Symphony**, **Obliv-C**, and **Wysteria** indicate the lines of code required (including the `main` harness), with (+$K$) denoting additional lines of library code. `?` means the program *should* be supported but we have no example, $*$ means a 2-party (rather than $N$-party) version should be supported, and empty indicates the program cannot be written in that language. Non-blank, non-comment lines were counted using `wc -l`.

| Program | # | Features | Symphony | Obliv-C | Wysteria | Description |
|---|---|---|---|---|---|---|
| `hamming` [14, 40] | 2 | | 16 | 44 | ? | Find the Hamming distance of two strings. |
| `edit` [39] | 2 | | 46 | 73 | ? | Find the edit distance, by dynamic programming, of two strings. |
| `bio-match` [7] | 2 | | 29 | 73 | ? | Compute the minimum Euclidean distance between a set of points (from $A$) and a single point (from $B$) in 2D space. |
| `db-analytics` [7] | 2 | | 94 | 99 | ? | Compute the mean and variance over the union and join of two databases. |
| `gcd` [21] | 2 | | 10 (+4) | 39 | 22 | GCD of two numbers via Euclid's algorithm. |
| `richest` [28] | $N$ | | 6 (+7) | | 34 | $N$-party variant of the Millionaire's. |
| `gps` [28] | $N$ | | 21 (+11) | | 42 | Compute the one-dimensional nearest neighbor for each of $N$ parties. |
| `auction` [28] | $N$ | | 13 (+4) | | 42 | Compute a second-price auction, revealing the second-highest bid to everyone and the highest bidder to auctioneer. |
| `median` [28] | 2 | `R` | 16 (+3) | ? | ? | Compute the mixed-mode (reactive) median of a set of numbers. |
| `intersect` [28] | 2 | `R` | 12 (+9) | ? | ? | Naive private set intersection over two sets. |
| `mmm` | $N$ | `P,$,R` | 29 (+5) | | | Use a comparison-based single-elimination tournament to find the richest of $N$ parties. |
| `committee` | $N$ | `P,D,$` | 26 (+45) | | | Elect a small committee of size $K < N$; useful for fair delegation from $N$ to $K$ parties. |
| `waksman` [34, 13] | $N$ | `$` | 167 (+67) | $*$ | | Securely shuffle of an array, using $N$ iterations of a Waksman permutation network. |
| `lwz` [23] | $N$ | `P,$,S` | 20 (+76) | | | Securely shuffle an array using $N$ reshares of a linear secret sharing scheme (LSSS). |
| `trivial-doram` [15, 6] | $N$ | | 23 | $*$ | ? | A library for Oblivious RAM, adapted to MPC from trivial client-server ORAM. |
| `shuffle-qs` [17] (Listing 4) | $N$ | `P,D,$,S,R` | 54 (+76) | | | Securely sort using Shuffle-Then-Sort with `lwz` as the underlying secure shuffle and QuickSort as the sorting algorithm. |

MOTION to add support for creating a `Party` object using existing TCP connections (rather than MOTION creating its own). This ensured that TCP connections were only established once for each pair of Symphony parties, rather than repeatedly by MOTION whenever a new `Party` object was created (i.e. on each decryption).

## 8 Experimental Evaluation

This section shows, through a series of experiments and case studies, that Symphony provides superior programming expressiveness and ergonomics compared to prior systems, while maintaining competitive performance.

## 8.1 Expressiveness and Ergonomics

We discuss SYMPHONY's expressiveness and ergonomics benefits based on our experience implementing 16 programs from the MPC literature. The programs are tabulated in Table 2. As points of comparison we consider if (or whether) versions of these programs could be implemented in Wysteria [28, 30] and/or Obliv-C [39], and whether they are $N$-party or 2-party programs. We categorize the SYMPHONY language features required to express the programs in the **Features** column: *Reactive MPC*, *Synchronized Randomness*, and *Delegation* (all described in Table 1 in Section 2), and *First-Class Party Sets* and *Resharing* (described in Section 3.2).

**Expressiveness** Of the sixteen programs, we believe that Obliv-C can express 9 and Wysteria can express 11. This is because both languages lack some needed features.

Obliv-C only supports two parties, ruling out fundamentally $N$-party programs. Obliv-C also lacks clean support for *Coordination*, as discussed in Section 2.2, but `mmm`, `committee`, `lwz`, and `shuffle-qs` require secure computation over *First-Class Party Sets* which are computed dynamically, based on values only available at runtime. Both the `mmm` and `committee` benchmarks were designed by us, and contain coordination patterns used to implement optimization techniques based on *adversary structures* [19].

Wysteria does not support *First-Class Party Sets*, *Delegation*, *Resharing*, or *Synchronized Randomness*. Wysteria does support party set values, but provides expression forms only for *creating* those values, not for *computing* with them. This precludes, for example, dynamically choosing a subset of parties to shuffle inputs, as used in LWZ. We discuss this serious limitation further in Appendix A.3. These limitations make it impossible to express `mmm`, `committee`, `waksman`, `lwz`, and `shuffle-qs`. Extending Wysteria with *Synchronized Randomness* would be straightforward, which would allow Wysteria to express `waksman`. The other programs still have barriers: they all require *First-Class Party Sets* and `committee`, `lwz`, and `shuffle-qs` additionally require *Delegation* or *Resharing*.

**Ergonomics** SYMPHONY can also provide ergonomics benefits even when a program could be expressed in another language. We use Wysteria as a point of comparison, since it also has coordination features.

Like SYMPHONY, and as described in Section 2.2, Wysteria coordinates parties by specifying which parties should execute in a given lexical scope via `let x =par(P)= e`, and where MPC should happen via `let x =sec(P)= e`. However, Wysteria does not allow recursive function calls in `sec` scope, which makes secure algorithms requiring bounded recursion awkward and inefficient.

**Listing 1** The `gcd` program of Table 2 as written in SYMPHONY.

```
def brec gcdr (a, b) =
  mux if (a == 0) then b
  else gcdr ((b % a), a)
def gcd = unroll gcdr (const 0) 93
```

**Listing 2** The gcd program of Table 2 as written in Wysteria.

```
1 let gcdr i (sa, sb) =
2   if i == 0 then
3     let ret =sec({!A,!B})= makesh 0 in
4     ret
5   else
6     let r =sec({!A,!B})=
7       let (a, b) = (combsh sa, combsh sb) in
8       (makesh (b % a), makesh a) in
9     let sr = gcdr (i - 1) r in
10    let ret =sec{!A,!B}=
11      let a = combsh sa in
12      if (a == 0) then sb
13      else sr
14    in ret
15 let gcd = gcdr 93
```

Consider the `gcd` function in Listings 1 and 2. This program computes the GCD of two encrypted integers among `{A,B}` in Symphony and Wysteria, respectively. The Symphony implementation is idiomatic and efficient, thanks to flexible first-class shares. By contrast, Wysteria's implementation requires multiple `sec` blocks due to secure computation both before the call to `gcdr` (`b % a` on line 8) and after (`if (a == 0)` on line 12). Note that there is no danger of a divide-by-zero on line 8—the encrypted operation `b % a` is total and evaluates to `b` when `a` is 0. The `gcd` function is recursive over encrypted values, so we must bound the recursion with a public upper bound (here, `93`) so that it can be expressed as a circuit. Symphony has a special `brec` keyword that facilitates this: `def brec gcdr a b` expands into `def gcdr gcdr a b`, with the `gcdr` parameter shadowing the recursive binding of the same name. The call to `unroll gcdr (const 0) 93` takes `gcdr` and "unrolls" it by performing self-application `93` times, using `const 0` as the base case.

Because Wysteria does not allow function calls within a `sec` scope, we are forced to enter and exit `sec` mode (lines 6,10) before and after each recursive call to `gcdr`. By comparison, Symphony's support for first-class shares allows naturally recursive algorithms like GCD to be expressed much more idiomatically under MPC.

### 8.2 Performance

Symphony's expressiveness does not place an undue burden on the implementation's ability to achieve good performance. We compared Symphony's performance with that of Obliv-C, a highly optimized MPC framework. Details of our experiments can be found in Appendices A.4 and A.5; here we summarize the results.

We ran on the first five programs in Table 2, which are well known and frequently referenced in the literature. We configured both Symphony and Obliv-C to use EMP's [36] 2-party garbled circuits implementation, to isolate language overhead from cryptography costs. On a simulated LAN under MPC, Symphony's running time was 1.15× that of Obliv-C. Without MPC, Symphony time is 2.4× that of Obliv-C. On a WAN (limited to 100 gbps and 50 ms RTTs), Symphony time was 0.85× that of Obliv-C. Examining these overheads, we find that one source is Symphony's support for $N > 2$ parties: party inclusion checks require the use of sets (implemented as balanced binary trees) rather than simple equality tests. The more significant overhead is unrelated to SYMPHONY itself: the language is implemented as an interpreter in Haskell, whereas Obliv-C is embedded within compiled C. Indeed, the sizes of garbled circuits generated by both frameworks are very similar. Thus, we would expect a narrower gap with a more production-quality implementation.

SYMPHONY's expressiveness also permits programmers to author algorithm-level optimization, directly and simply. For example, the `median` and `shuffle-qs` programs leverage *Reactive MPC* to perform certain comparison operations in the clear, dramatically improving performance over a monolithic protocol [22, 29]. While Reactive MPC is available in some languages, the combination of features needed to express the `lwz` secure shuffle is unique to SYMPHONY—no other language can express it. Compared to `waksman`, another secure shuffle algorithm, `lwz` offers a substantial performance benefit because it requires *no computation under cryptography*, just re-sharing and comparisons/shuffling in the clear. The result is dramatically faster running times for all input sizes.

## 9 Conclusion

We have presented SYMPHONY, a new language for MPC with strong support for *coordination*. SYMPHONY's scoped *par* blocks, first-class party sets, and first-class shares—which support reactive computation, delegation, and resharing—provide unparalleled expressiveness which along with SYMPHONY's generalized SIMD semantics make coordination programming less error prone. We formalized core SYMPHONY and proved that the intuitive, single-threaded interpretation of a program coincides with its actual distributed semantics. Our prototype implementation exhibits performance competitive with existing systems, while uniquely enabling optimizations.

## A Implementation, Experiments, and Case Studies

### A.1 Implementation

This section provides additional information about the SYMPHONY Haskell interpreter (extending Section 7). We highlight three interesting features: MPC over algebraic types, synchronized randomness, and the standard library.

**MPC over Algebraic Types** SYMPHONY generalizes $\lambda$-SYMPHONY's `share`, `mux`, `case`, and `reveal` by allowing arbitrary algebraic types as arguments. It also adds another expression, `mux-case`, for case analysis on encrypted (shared) values. The `share`, `mux-case`, and `reveal` expressions on pairs are generally unsurprising. Sharing a product (pair) is implemented by recursively sharing, component-wise.

$$\texttt{share}[\phi, \tau_1 \times \tau_2 : P \rightarrow Q]\,(a,\ b) \;=\; (\texttt{share}[\phi, \tau_1 : P \rightarrow Q]\ a,\ \texttt{share}[\phi, \tau_2 : P \rightarrow Q]\ b)$$

The `mux-case` operation on encrypted sums is identical to `case`, and the `reveal` operation is works similarly to `share`.

We represent sums (variants) as tagged pairs. The λ-SYMPHONY value $\iota_0\ v$ is represented as $\texttt{sum}\langle\texttt{true}, v, \texttt{default}\rangle$ and the value $\iota_1\ v$ is represented as $\texttt{sum}\langle\texttt{false}, \texttt{default}, v\rangle$. The value `default` is a placeholder which will be replaced by a default value of the appropriate type if the sum value is shared. Sharing a sum value is implemented, as with pairs, by recursively sharing, component-wise.

$$\texttt{share}[\phi, \tau_1 + \tau_2 : P \to Q]\ \texttt{sum}\langle b, v_1, v_2\rangle$$
$$\equiv \texttt{sum}\langle\texttt{share}[\phi, \mathbb{B} : P \to Q]\ b, \texttt{share}[\phi, \tau_1 : P \to Q]\ v_1, \texttt{share}[\phi, \tau_2 : P \to Q]\ v_2\rangle$$

To share a `default` value, $\texttt{share}[\phi, \tau : P \to Q]\ \texttt{default}$, we simply share instead the default value of the appropriate type $\tau$. It is important for `mux-case` that the type $\tau$ be a monoid and that the default value is the identity. For example, when $\tau$ is `bool` the identity is `false` and the (monoidal) operation is `xor`. Likewise, when $\tau$ is `nat` the identity is 0 and the operation is +. The `reveal` operation on sum values works similarly.

The $\texttt{mux-case}\ \texttt{sum}\langle b, v_0, v_1\rangle\ \{\theta_1 \to e_1; \ldots; \theta_n \to e_n\}$ expression is the most interesting one we have to consider. The tag, $b$, is *encrypted* and so we cannot inspect it to determine if it matches a left or right injection pattern. So, what do we do? The `mux-case` proceeds by first filtering out any patterns $\theta_1, \ldots, \theta_n$ which are not either a left injection or a right injection pattern. Then, it *maps* over the remaining patterns $\theta_k \equiv \iota_i x$[5] by producing $\texttt{mux}\ b\ \texttt{then}\ \texttt{default}\ \texttt{else}\ v_0'$ when $i$ is 0 and $\texttt{mux}\ b\ \texttt{then}\ v_1'\ \texttt{else}\ \texttt{default}$ when $i$ is 1 where $v_i'$ is the result of evaluating $e_k$ under the environment ($\gamma$) extended with $[x \mapsto v_i]$. At this point, we `add` together the list $v_1, \ldots, v_m$ (where $m \leq n$ corresponds to the number of patterns in $\theta_1, \ldots, \theta_n$ which were either a left or right injection) of values just produced. The `add` procedure is the monoidal operation associated with the type, $\tau$, of the values $v_1, \ldots, v_m$.

Consider the standard encoding of booleans as a sum of units: $\texttt{bool} := \texttt{unit} + \texttt{unit}$. We will take $\texttt{sum}\langle\texttt{true}, \bullet, \bullet\rangle$ to be the encoding of `true` and symmetrically for `false`. We would hope that the obvious encoding of `mux` in terms of `mux-case` would work: $\texttt{mux}\ b\ \texttt{then}\ e_1\ \texttt{else}\ e_2 := \texttt{mux-case}\ \texttt{sum}\langle b, \bullet, \bullet\rangle\ \{\iota_{\texttt{true}}\ \bullet \to e_1;\ \iota_{\texttt{false}}\ \bullet \to e_2\}$. Indeed, it does. In the following calculation, let $v_i$ denote the result of evaluating $e_i$. Let $\tau$ be the type of $v_1$ and $v_2$, $\texttt{id}_\tau$ denote the monoidal identity on $\tau$ and $\texttt{add}_\tau$ denote the monoidal operation. Then, we have:

$$\begin{aligned}
&\texttt{mux-case}\ \texttt{sum}\langle\texttt{true}, \bullet, \bullet\rangle\ \{\iota_0\ \bullet \to e_1; \iota_1\ \bullet \to e_2\} \\
&\equiv \texttt{add}_\tau\ (\texttt{mux}\ \texttt{true}\ \texttt{then}\ v_1\ \texttt{else}\ \texttt{id}_\tau) \\
&\qquad\quad (\texttt{mux}\ \texttt{true}\ \texttt{then}\ \texttt{id}_\tau\ \texttt{else}\ v_2) \\
&\equiv \texttt{add}_\tau\ v_1\ \texttt{id}_\tau \\
&\equiv v_1
\end{aligned}$$

MPC over algebraic types is an interesting problem, and there's a lot of unsolved problems in this space. We recommend that interested readers consult Oblivious Algebraic Data Types by Ye and Delaware [38] for a more complete treatment.

[5] We are ignoring nested patterns for simplicity. The procedure described here is straightforward to extend.

**Synchronized Randomness** Symphony provides a convenient way to generate randomness across multiple parties in parallel, ensuring that all parties receive the same random value. This functionality can be implemented as a library using only access to local randomness, as shown in the Symphony code below.

```
def randomSend P n =
  let rec sum' = fun Q ->
    case Q
    { {}          -> 0
    ; { p } \/ Q' ->
      send [nat : p -> P]
        (par { p } rand { p } nat) + (sum' Q')
    }
  in
  let sum = sum' P in
  sum % n
```

This code will generate a random value on each party in `P` and sum them all together, modulo `n`, to generate a random natural number in the range $0..n-1$. This works, but it is inefficient because it requires communication between all the parties each time they wish to generate a synchronized random number.[6]

The primitive expression provided by Symphony, `rand P` $\mu$ (where $\mu$ is a base type like `nat`), provides the same functionality but does so more efficiently. It establishes a shared seed among the parties `P` using the procedure above the first time they request a synchronized random number. Thereafter, the parties use a local, cryptographically secure pseudo-random generator which requires no communication.

**The Standard Library** The standard library shipped with Symphony has many of the usual fixings of functional programming languages such as libraries for options; eliminators (folds) over various algebraic types (nats, options, lists, maps, etc.); higher order functions (flip, compose, curry, uncurry, etc.). However, it also has some unusual fixings, such as libraries for coordination, bounded recursion, and utilities for synchronized randomness. Next, we highlight some interesting, representative functions.

[6] Readers may recognize that the random number generated by this procedure is slightly biased. When we use this procedure to establish a seed (described below) it is unbiased because the seed is a multiple of the word size of the machine (128 bits in practice).

**Listing 3** Selected functions from the standard library of SYMPHONY.

```
def bundleUpWith f P = case P
  { {}          -> <<>>
  ; { p } \/ P' -> << p | par { p } f p >> ++ bundleUpWith f P'
  }

def unroll f init n =
  if n == 0n then init
  else f (unroll f init (n - 1n))

def randNat P               = rand P nat
def randMaxNat P max        = randMax P nat max
def randRangeNat P min max = min + (randMaxNat P (max - min))

def shuffle P a =
  let n = size a in
  let shuffleRec = fun [shuffleRec] i ->
    if i < n - 1n then
      let j = randRangeNat P i n in
      let _ = swap a i j in
      shuffleRec (i + 1n)
    else ()
  in if n >= 2n then shuffleRec 0n else ()

def permutation P n =
  let a = upTo n in
  let _ = shuffle P a in
  a
```

The coordination library primarily addresses common operations on principal sets, bundles, and their interaction. For example, the `bundleUpWith` function in Listing 3 takes a function `f` and a set of parties,`P`, and creates a bundle among the parties `P` by calling `f` locally on each party `p` in `P`. This function is useful for eliminating the boilerplate associated with the common pattern of reading an input from each party.

```
def readInputs P = bundleUpWith (fun _ -> read nat) P
```

The bounded recursion library in SYMPHONY provides a critical function, `unroll`, which was used and briefly described in Listing 1. The full definition of `unroll` appears in Listing 3. As described in Section 8.1, the `unroll f init n` function is used in combination with the `brec` keyword to finitely "unroll" a function, `f`, `n` times before eventually bottoming out with a call to `init`. To illustrate, consider the `gcd` and `gcdr` functions from Section 8.1.

```
def brec gcdr (a, b) = mux if (a == 0) then b else gcdr ((b % a), a)
def gcd = unroll gcdr (const 0) 93
```

The signature `def brec` $f\ x_1 \dots x_n$ desugars into the signature `def` $f\ f\ x_1 \dots x_n$ so that any recursive call in the body of $f$ now instead refers to the formal parameter $f$ rather than the function being defined. By shadowing the recursive binding, we transparently turn any recursive function into one that can be used for *bounded* recursion. Here's what that looks like for `gcdr`:

```
def gcdr gcdr (a, b) = mux if (a == 0) then b else gcdr ((b % a), a)
```

Finally, we can compute the finite unrollings for 0 and 2 to see how `unroll` works in conjunction with functions desugared using `brec`.

```
def gcd = unroll gcdr (const 0) 0
        = if 0 == 0 then (const 0) else gcdr (unroll gcdr (const 0) (0-1)
  ↪ )
        = const 0
```

```
def gcd = unroll gcdr (const 0) 2
        = if 2 == 0 then (const 0) else gcdr (unroll gcdr (const 0) (2-1)
  ↪ )
        = gcdr (unroll gcdr (const 0) 1)
        = gcdr (if 1 == 0 then (const 0) else gcdr (unroll gcdr (const 0)
  ↪  (1-1)))
        = gcdr (gcdr (unroll gcdr (const 0) 0))
        = gcdr (gcdr (const 0))
```

**FFI and Resource Management** The Haskell interpreter for Symphony implements MPC using the Symphony runtime library. The Symphony runtime library implements MPC using the EMP and MOTION libraries. Each of these layers (Symphony interpreter, Symphony runtime, and EMP/MOTION) interact via FFI, using opaque foreign pointers to represent values in the next layer.

In Haskell, the raw foreign pointers are wrapped with the ForeignPtr type, which executes an associated *finalizer* (i.e. a Haskell function) on the raw foreign pointer when it is garbage collected. We use this finalizer to call (via FFI) an appropriate destructor function which is provided by Symphony runtime.

In Rust, the raw foreign pointers are wrapped with `newtype`-style Rust structures. These Rust structures contain an explicit implementation of the Drop trait, calling the `drop` function on the structure when it is dropped. This is analogous to the `ForeignPtr` in Haskell, except that Rust can insert calls to `drop` statically rather than relying on garbage collection. Just as in Haskell, the `drop` function calls (via FFI) an appropriate destructor function, which is provided by EMP and MOTION.

Finally, we implement FFI interfaces for both EMP and MOTION which expose constructor and destructor functions which perform heap allocation and deallocation of the requisite C++ objects.

Putting all this together, when an encrypted value is garbage collected by Haskell the `ForeignPtr` will call the appropriate destructor defined by Symphony runtime. Then, that destructor function will drop the value which will call the appropriate destructor defined by EMP or MOTION. Finally, the destructor of EMP or MOTION exposed by the FFI will free the C++ object using the `delete` keyword.

These approaches to integrating software written in different languages are largely standard. We chose to implement the enhancements in Symphony runtime as a separate library to optimize performance. We chose Rust specifically because it has excellent libraries, tooling, and documentation. The Symphony runtime is written in idiomatic Rust, meaning that it may serve as an artifact of independent interest for researchers who need access to MPC protoocols with support for delegation, resharing, and reactive MPC.

### A.2 Benchmark Programs

Here we describe some of the programs we have implemented in Symphony, tabulated in Table 2, in more detail. The language *features* required to implement each program are briefly discussed in Section 8.1.

The implementation of all the programs listed here, as well as the templates used for benchmarking, can be found in Symphony's repository: https://github.com/plum-umd/symphony-lang.

**hamming**, **edit**, **bio-match**, **db-analytics**, and **gcd**, are 2-party programs, fully described in Table 2. We implemented these programs in Symphony and Obliv-C, which is limited to 2-party MPC. We believe they could be implemented in Wysteria.

**richest**, **gps**, **auction**, **median**, and **intersect** are also fully described in Table 2. They were previously implemented in Wysteria and we back-ported them to Symphony. **richest**, **gps**, and **auction** support $N$ parties but otherwise require no special coordination. **median** and **intersect** support 2 parties and use *Reactive MPC* (referred to as "mixed-mode" by Wysteria) to improve their efficiency; we believe these could be implemented in Obliv-C.

**mmm** orchestrates a single-elimination tournament over $N$ parties. Each match in the tournament is a 2-player secure computation, with the winner moving on to the next round. The program requires *First-Class Party Sets* because the participants in each match are dynamically assigned from the set of remaining players. Likewise, the set of remaining players is determined dynamically according to outcome of the matches in the previous round. It requires *Synchronized Randomness* to determine the participants in each match. Finally, it requires *Reactive MPC* to decrypt the winners in each round so that they may be coordinated in the following round.

**committee** performs an arbitrary secure computation among $N$ parties by selecting a random committee of size $K < N$ and delegating the secure computation to them. The program requires *First-Class Party Sets* and *Synchronized Randomness* because the committee is computed dynamically according to $K$ and synchronized random numbers generated by the $N$ parties. It requires *Delegation* because the parties outside the committee encrypt their input and send it to the committee but do not participate in the secure computation.

**waksman** is a protocol for securely shuffling an array of elements among $N$ parties without revealing the permutation to any of them. The Waksman protocol implements a secure shuffle via repeated application of a classic Waksman *permutation network* [34]. A permutation network repeatedly and conditionally swaps two list elements at a time until the list is fully shuffled; each swap can be implemented from Boolean gates. One subtlety of a permutation network is that one must choose the *control bits* of the network, which is an input that dictates which of the swap gates should indeed swap their input. To fully hide the shuffle from all parties, each party secretly chooses their own control bits and programs one of the sequence of $|P|$ networks. Thus, the full shuffle requires $|P|$ networks, and the implementation requires some coordination: the programmer must prescribe that each party will, one-by-one, program a network. This protocol requires *Synchronized Randomness* to perform (non-secure) shuffles as a subroutine.

**lwz** is also a secure shuffle protocol, originally due to Laur, Willemson, and Zhang [23]. They showed how to implement a highly efficient secure shuffle among parties $Q$ which is resilient to $T$ corruptions, using a linear secret sharing scheme. The LWZ protocol implements a secure shuffle by repeatedly shuffling the elements among *committees*, which are strict subsets of the parties of size $|Q|-T$. A committee is given shares of the input list and agrees on a random permutation, $\pi$; locally permutes its shares according to $\pi$; and then constructs new shares for the next committee. This process repeats until each set of $T$ parties has been excluded from some committee, which is sufficient to hide the global shuffle.[7] In addition to *Synchronized Randomness* **lwz** requires *First-Class Party Sets* and *Resharing* to compute all the party sets of size $N-T$ and reshare the array of elements to and from these subsets. As far as we are aware, no prior MPC language can directly express **lwz**.

**trivial ORAM** Trivial ORAM [35] allows secret shares to be used as addresses by simulating random access. The simulation performs a linear scan of the underlying array for every access (read or write). Multiplexors are used to choose which index to operate on. Trivial ORAM is a basic primitive which is used to construct more efficient ORAM implementations such as Circuit ORAM.

[7] It is also necessary for security that $|Q|-T > T$ (equivalently, $|Q| \geq 2T+1$) or else a corrupt committee of size $|Q|-T$ can collude to reveal the secret elements being shuffled.

**Listing 4** A secure, $N$-party sorting procedure written in Symphony. Uses the Shuffle-Then-Sort paradigm [17] with LWZ as the underlying shuffle. Each party in $\{A, B, C, D, E\}$ contributes an array of integers, which are concatenated together and then securely shuffled and sorted by the parties in Q.

```
1  party A B C D E
2
3  -- read input at p, secret-share to all in Q
4  def readShare Q p = par ({ p } \/ Q)
5    let i = par { p } read (array int) from "lwz.txt" in -- file local to
   ↪ each p
6    share [gmw, array int : { p } -> Q] i
7
8  def delegateShares P Q =
9    map (readShare Q) (psetToList P)
10
11 def shuffleWith Q S sharesQ = par (Q \/ S)
12   let sharesS = share [gmw, array int : Q -> S] sharesQ in
13   share [gmw, array int : S -> Q] (shuffle S sharesS)
14
15 def lwz Q sharesQ =
16   let t = 1 in
17   foldr (shuffleWith Q) sharesQ (subsets Q ((psetSize Q) - t))
18
19 def revealLte Q x y = reveal [gmw, bool : Q -> Q] x <= y
20
21 def secureSort Q sharesList =
22   let sharesQ  = par Q (arrayConcat sharesList) in
23   let shuffled = lwz Q sharesQ in
24   let sorted   = quickSort (revealLte Q) shuffled
25
26 def main () = par {A,B,C,D,E}
27   let Q = {A,B,C} in
28   let sharesList = delegateShares {A,B,C,D,E} Q in
29   let sorted     = secureSort Q sharesList in
30   ...
```

**shuffle-qs** is the most sophisticated program, using both **committee** and **lwz** as subroutines. It implements a secure sorting procedure over $N$ parties by choosing a committee of size $K > 2$, shuffling the elements among the committee using **lwz**, and then sorting the elements with QuickSort by revealing the result of each comparison.

The **shuffle-qs** program as written in Symphony is shown in Listing 4. As is standard, all parties run the same program, starting at `main`. While the shuffling and sorting code works for arbitrary numbers of parties, `main` is specialized to those parties declared at the top, named A–E. When `par {A,B,C,D,E}` ... is reached on line 26, only the listed parties execute the subsequent code .... On line 27, `Q` is bound to the set `{A,B,C}` and is then passed as the second argument to `delegateShares` on line 28, and as the first to `secureSort` on line 29.

In `readShare Q p`, party `p` reads an array of integers from local file `lwz.txt` (line 5), and then creates a share among parties in `Q` (in Symphony a share of an array is an array of shares). The `share` operation requires all parties in $\{$`p`$\} \cup$ `Q` to be present (ensured by the `par` on line 4) so that `p` can transmit to each party in `Q` its share and know they are ready to receive it—note that `p` may or may not be a member of `Q`. The `delegateShares` function calls `readShare Q` for each party `p` ∈ `P`, with the goal of *delegating* the subsequent computation to those parties in `Q`.

On lines 12–13 the shares among parties `Q` are reshared to be among parties `S`, and then reshared back to `Q` once shuffled by `S`. The `foldr` on line 17 invokes `shuffleWith` on each subset `S` of `Q` (computed by `subsets`), which has size $|Q| - T$. In turn, this function reshares `sharesQ` among those parties in `S`, which invoke `shuffle` to permute its values, and then reshare the result back. Within `shuffle`, the parties `S` agree on a seed for a PRNG that they use as the basis for the shuffle, ensuring they compute the same permutation. Laur, Willemson, and Zhang proved that if each subset `S` has size $|P| - T$, then nothing can be learned about the order of the shuffled elements unless $n > T$ parties collude. In Listing 4, we specify $T = 1$ on line 16.

Once line 23 completes, `shuffled` contains a shuffled, secret-shared array. Line 24 then `quicksorts` this array, using `revealLte Q` as the comparison function. This function uses SYMPHONY's `reveal` construct to convert the shares among `P` to plaintext values replicated among `Q`.

This program requires *First-Class Party Sets*, *Delegation*, *Synchronized Randomness*, and *Resharing* by virtue of relying on **committee** and **lwz**. It requires *Reactive MPC* because each comparison in QuickSort is decrypted before securely computing the next comparison.

### A.3 Committee Election in Wysteria

Wysteria does not support first-class party sets due to a lack of support for *computing* on them. We can see this in an attempt to port the `elect` function in Listing 5 from SYMPHONY to Wysteria, which ultimately proves impossible. This limitation impacts Wysteria's ability to express coordination protocols compositionally.

**Listing 5** A simple SYMPHONY function to elect a subset of $<= k$ parties from among `P`. A more sophisticated version of this function is used in the `committee` program (see Table 2).

```
def elect P k =
  if k == 0 then { }
  else case P
    { { }          -> { }
    ; { p } \/ P' -> { p } \/ (elect P' (k - 1))
    }
```

The `elect P k` function "elects" a subset of at most `k` parties from the set `P`. A more featureful implementation might use a voting procedure or synchronized randomness to determine the subset, but this simple version will suffice for our purposes. Let's consider how we might express this function in Wysteria.

**Attempt 1** The most natural way to express `elect` is through a signature that matches that of SYMPHONY.

```
let elect (P : ps{true}) (k : nat) : ps{subeq P} = ??
```

Here we see a fairly typical OCaml-like signature with some type annotations. The type system of Wysteria is a refinement type system, but restricted so that refinements can only be placed on the party set type, `ps`. The set of refinements contains the standard propositional refinements `true` and conjunction (`and`), as well as special refinements `singl`, `subeq x`, and `eq x`. The `singl` refinement is satisfied by sets, $P$, that are singletons: $|P| = 1$. The `subeq Q` (`eq Q`) refinement is satisfied by party sets, $P$, that are a subset of (equal to) $Q$: $P \subseteq Q$.

In this signature for `elect`, we see that `P` has type `ps{true}` indicating that it is an unrefined party set. In addition, we see that `elect` should return a party set which is a subset of `P`. The fact that `P` is considered bound in the return type is a hallmark of dependent and refinement type systems.

Unfortunately, this attempt is dead on arrival. We cannot implement this signature because Wysteria simply has no expression form that corresponds to elimination (i.e. computation on) party sets. In our next attempt, we will try to circumvent this restriction by *encoding* party sets as arrays and computing over arrays instead.

**Attempt 2** In this attempt, we will encode a party set `P` as an array containing singleton sets over elements belonging to `P` according to the type `array ps{singl and subeq P}`. This approach shows much more promise:

```
let elect (P  : ps{true})
          (Pe : array ps{singl and subeq P})
          (k  : nat)
          : array ps{singl and subeq P} =
  let ret = array [ k ] of (select Pe[0]) in
  let rec elect_loop i =
    if i < k then
      let _ = update ret[i] <- select Pe[i] in
      elect_loop (i + 1)
      else ()
  in
  let _ = elect_loop 0 in
  ret
```

This function loops through the first `k` elements of `Pe` and assigns them to a new array, `ret`, before returning `ret`. Contrary to our last attempt, the parameter `P` only exists in this signature to refine the type of `Pe` and the return type. All the computation is being performed over our encoding of party sets.

This version of `elect` is well-typed, and so we can try using it in a simple example.

```
let foo (Q : ps{true}) (Qe : array ps{singl and subeq Q}) : ... = ...

let P  = { !A, !B, !C } in
let Pe = array [ 3 ] of { !A } in
let _  = update Pe[1] <- { !B } in
let _  = update Pe[2] <- { !C } in
let Ce = elect P Pe 2 in
let ret =par(??)=
  foo ?? Ce
in ret
```

This little driver program declares `P` to contain the parties `A`, `B`, and `C` which are encoded in `Pe`. It uses the `elect` function above to produce an encoded committee, `Ce`, of 2 parties. At this point we would like just the committee members to run `foo`, but to do so we must fill the holes (`??`) with the actual party set represented by the encoding `Ce`. To do so, we define a `decode` function.

```
let decode (P : ps{true})
           (Pe : array ps{singl and subeq P})
           : ps{subeq P} =
  let n = length Pe in
  let rec decode_loop i acc =
    if i < n then
      let p = select Pe[i] in
      decode_loop (i + 1) (acc \/ p)
    else acc
  in decode_loop 0 { }
```

The `decode` function iterates over `Pe` and takes the union of all its elements. Since each of these elements is labeled as being a subset of `P`, Wysteria is able to deduce that the union of all the elements is also a subset of `P` as indicated by the return type `ps{subeq P}`.[8] Let's try using our `decode` function to fill in those holes from our earlier example.

```
let foo (Q : ps{true}) (Qe : array ps{singl and subeq Q}) : ... = ...

let P  = { !A, !B, !C } in
let Pe = array [ 3 ] of { !A } in
let _  = update Pe[1] <- { !B } in
let _  = update Pe[2] <- { !C } in
let Ce = elect P Pe 2 in
let C  = decode P Ce in
let ret =par(C)=
  foo C Ce
in ret
```

We can now enter an execution context containing only the parties `C` using `par(C)` mode. Unfortunately, the type system rejects this program at the call to `foo`. Why? Because Wysteria is unable to deduce that `Ce` has type `array ps{singl and subeq C}` as required by `foo`'s signature. Instead, the type of `Ce` is `array ps{singl and subeq P}`. We could imagine using a different signature for `foo`, but it is important for compositionality that `foo` take an encoded party set. Doing so enables `foo` to leverage any other functions that dynamically compute party sets according to our encoding. For example, `foo` might wish to call `elect` on its argument.

Of course, the critical issue is that the return type of `elect` is not precise enough, thereby preventing the encoded party set returned by `elect` from being passed to other functions that need to compute over party sets. At this point we are stuck.

[8] Wysteria does not support the `length` primitive on arrays that we used to compute `n`. This could be easily added, and allows us to avoid cluttering up our signatures with array lengths.

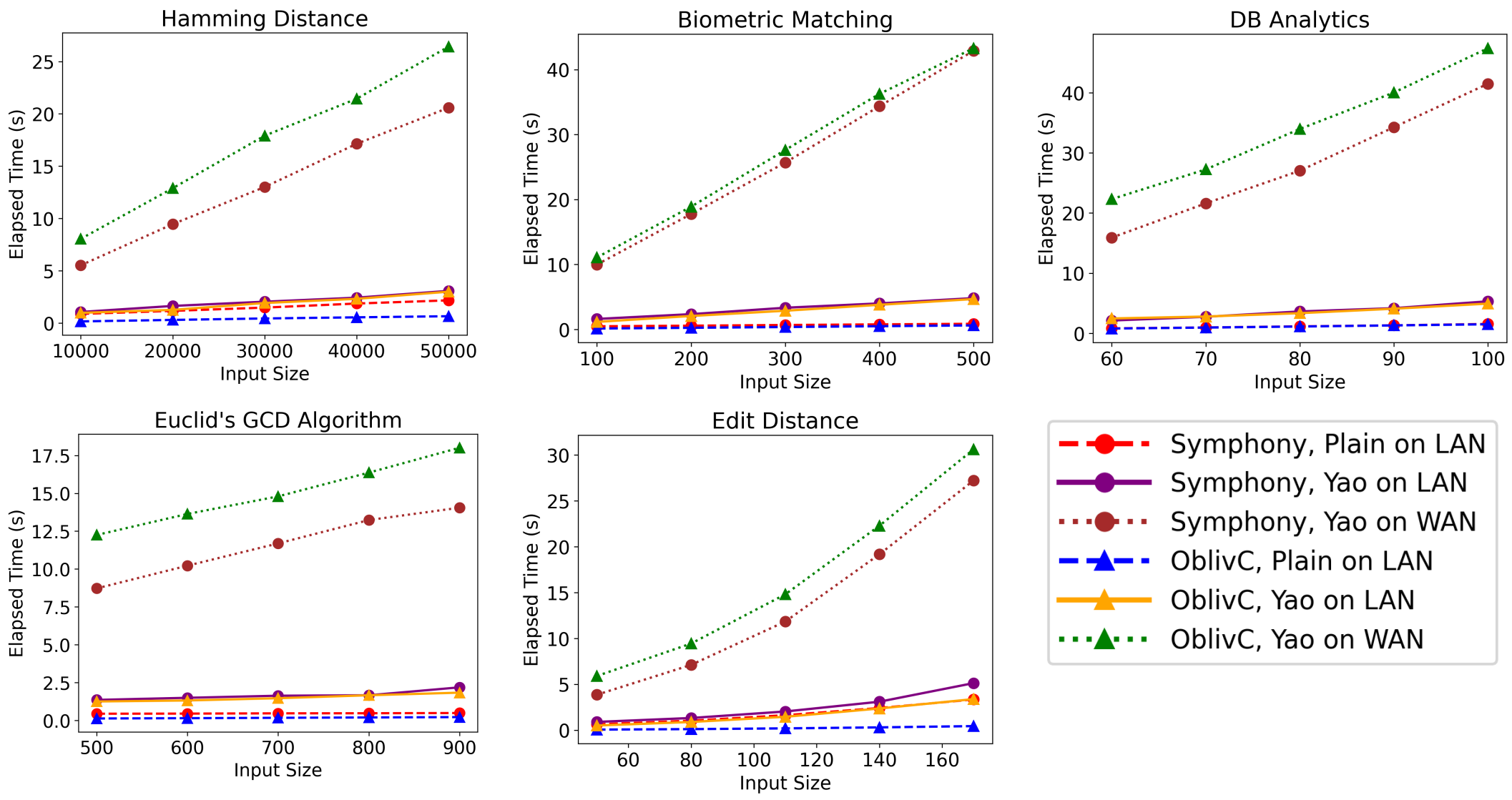

**Figure 7** End-to-end execution time of 5 programs, averaged over five samples (lower is better). **LAN** is a simulated 1 gbps connection with no delay. **WAN** is a simulated 100 mbps connection with a 50 ms RTT latency. **Yao** and **Plain** protocols use EMP's `sh2pc` (semi-honest, two-party) and `plain` protocols respectively. Symphony uses EMP's Integer interface. Obliv-C uses EMP's Bit interface (compiles integer operations to circuits). Input sizes for all the benchmarks indicate the length of the list(s) provided as input, except for `gcd-gc` where the input size indicates the number of iterations of the GCD algorithm.

Wysteria does not offer any other facilities to circumvent the lack of an elimination form on party sets.

### A.4 Performance vs Obliv-C

We compared Symphony's performance, in terms of running time and gate counts, against that of Obliv-C on the same programs. The results are summarized in Section 8.2; this section provides details of the experiments and results.

**Experimental Setup** For fair comparison, both Symphony and Obliv-C were configured to use EMP [36] as their MPC backend; we extended Obliv-C to use EMP via its callback interface. We used both Symphony and Obliv-C to implement a benchmark suite of five programs: `hamming`, `edit-dist`, `bio-match`, `db-analytics`, and `gcd`. See Table 2 for a description of these programs.

Experiments were run on a 2019 MacBook Pro with a 2.8 GHz Quad-Core Intel Core i7 and 16 GB of RAM (OSX 11.3.1). The Obliv-C compiler is an extension of GCC 5.5.0, and all benchmarks were compiled with `-O3` optimizations. Experiments were run on two simulated networks: a LAN (1 gbps bandwidth, <1 ms RTT latency) and a WAN (100Mbps bandwidth, 50 ms RTT latency). All experiments use 32-bit integers, except for `gcd-gc` which uses 64-bit integers. Reported execution times measure the end-to-end execution time of party *A* and were averaged over five samples.

**Running time** Figure 7 plots the end-to-end execution time of Symphony and of Obliv-C on the benchmarks. On LAN under MPC (Yao), Symphony's running time is 1.15× that of Obliv-C (per the geometric mean). Without MPC, Symphony time is 2.4× that of Obliv-C. On WAN under MPC, Symphony time is 0.85× that of Obliv-C. The maximum slowdown occurs in `edit-dist`, which uses dynamic programming and for Obliv-C is heavily optimized by GCC.

There are a two primary sources for Symphony's overhead: First, Symphony supports arbitrary numbers of parties while Obliv-C supports only two. This is significant because Symphony performs frequent runtime checks on the parties in scope. Since Symphony supports an arbitrary number of possible parties, we represent the parties in scope as a set (implemented by a balanced tree data structure). Thus checks on principals are implemented by set operations. We could improve the efficiency of these runtime checks by implementing them using a bitset instead of a balanced tree. Obliv-C also performs certain checks on parties but, since only two parties are supported, these are implemented as simple integer equality checks.

Second, Symphony is interpreted but Obliv-C is compiled. Interpretation imposes overhead, especially for programs involving loops. For example, a simple stress test which sums 1 million integers (in the clear) on a single party shows that Symphony takes about 6 seconds where Obliv-C takes about 100 milliseconds. This stress test executes no runtime checks imposed by $\lambda$-Symphony, which suggests that the overhead is due to interpretation.

Since both Symphony and Obliv-C are synchronous (i.e. they block when reading from the network), each non-local MPC operation imposes a RTT delay on the real execution time. If the implementations were asynchronous instead, the MPC operations and interpretation would execute in parallel. Instead of an additive delay, real execution time between non-local MPC operations would be the maximum of the interpretation time and RTT. For all but the fastest LAN networks, the RTT is $> \ 5\,ms$. We conjecture that the interpretive overhead of Symphony is small enough that it is dominated entirely by the network latency for most deployments. If that is the case, real execution time between asynchronous Symphony and Obliv-C would be indistinguishable.

Comparing the LAN and WAN benchmarks confirms that the language overhead imposed by Symphony is dominated by the time it takes to perform network communication during a WAN deployment of MPC. We believe Symphony is faster than Obliv-C in the WAN setting due to its use of the EMP Integer interface, which uses the network more efficiently than the Bit interface used by Obliv-C's callback mechanism, and consequently the EMP backend for Obliv-C.

**Generated circuit sizes** As a second experiment, we instrumented the EMP backend to count the number of utilized AND and XOR gates. Counting gates is primarily a sanity check that ensures Symphony is not erroneously introducing large numbers of unneeded gates. Table 3 tabulates the number of AND and XOR gates generated by Symphony and by Obliv-C. The gate counts generated by Symphony and Obliv-C are very similar, with differences caused by using the EMP Integer interface vs Obliv-C compiling to the Bit interface (as required by its callback mechanism. The optimizations performed by EMP's circuit compiler and Obliv-C's circuit compiler are similar, but not identical.

**Table 3** Gate counts (AND and XOR) of select benchmark programs. **Input Size** for Hamming Dist., Bio. Matching, DB Analytics, and Edit Dist. is the length of the input lists. For GCD, it is the maximum number of GCD iterations. Gate counts were collected by modifying EMP to record AND or XOR gate execution. Symphony uses EMP's Integer interface where applicable, OblivC uses EMP's Bit interface (compiling integer operations to circuits).

| | | OblivC | | Symphony | | Δ (OblivC - Symphony) | |
|---|---|---|---|---|---|---|---|
| Benchmark | Input Size | AND Gates | XOR Gates | AND Gates | XOR Gates | AND Gates | XOR Gates |
| Hamming Dist. | 10000 | 1249875 | 3159595 | 950000 | 2550000 | 299875 | 609595 |
| | 20000 | 2499875 | 6319595 | 1900000 | 5100000 | 599875 | 1219595 |
| | 30000 | 3749875 | 9479595 | 2850000 | 7650000 | 899875 | 1829595 |
| | 40000 | 4999875 | 12639595 | 3800000 | 10200000 | 1199875 | 2439595 |
| | 50000 | 6249875 | 15799595 | 4750000 | 12750000 | 1499875 | 3049595 |
| Bio. Matching | 100 | 2617868 | 6353496 | 2675500 | 8007000 | -57632 | -1653504 |
| | 200 | 5235768 | 12706996 | 5351000 | 16014000 | -115232 | -3307004 |
| | 300 | 7853668 | 19060496 | 8026500 | 24021000 | -172832 | -4960504 |
| | 400 | 10471568 | 25413996 | 10702000 | 32028000 | -230432 | -6614004 |
| | 500 | 13089468 | 31767496 | 13377500 | 40035000 | -288032 | -8267504 |
| DB Analytics | 60 | 4609304 | 9569553 | 4732457 | 10425422 | -123153 | -855869 |
| | 70 | 6246968 | 12970159 | 6413597 | 14128242 | -166629 | -1158083 |
| | 80 | 8133020 | 16886701 | 8349937 | 18393262 | -216917 | -1506561 |
| | 90 | 10268008 | 21320663 | 10541477 | 23220482 | -273469 | -1899819 |
| | 100 | 12651370 | 26270229 | 12988217 | 28609902 | -336847 | -2339673 |
| GCD | 500 | 2360091 | 6735821 | 2302192 | 6910016 | 57899 | -174195 |
| | 600 | 2832991 | 8085621 | 2762592 | 8291916 | 70399 | -206295 |
| | 700 | 3305891 | 9435421 | 3222992 | 9673816 | 82899 | -238395 |
| | 800 | 3778791 | 10785221 | 3683392 | 11055716 | 95399 | -270495 |
| | 900 | 4251691 | 12135021 | 4143792 | 12437616 | 107899 | -302595 |
| Edit Dist. | 50 | 780882 | 1704539 | 637372 | 1779607 | 143510 | -75068 |
| | 80 | 2003142 | 4378936 | 1631872 | 4556407 | 371270 | -177471 |
| | 110 | 3790602 | 8291824 | 3085372 | 8614807 | 705230 | -322983 |
| | 140 | 6143262 | 13443100 | 4997872 | 13954807 | 1145390 | -511707 |
| | 170 | 9061122 | 19832759 | 7369372 | 20576407 | 1691750 | -743648 |

Overall, our experiments indicate that the language design itself does not impose significant overhead on either end-to-end execution time or generated circuit sizes. We leave a more sophisticated implementation which leverages compilation and compiler optimizations to future work.

## A.5 Security vs. Performance for Secure-Shuffle

The Waksman and LWZ protocols present different tradeoffs in terms of security and performance.

Given an input list of $n$ integers of bitwidth $w$, a Waksman permutation network is a recursive algorithm requiring $O(w \cdot n \log n)$ Boolean gates. Since we must repeat the network $|P|$ times, we require $O(|P| \cdot w \cdot n \log n)$ gates total. The network's circuit depth grows with $O(\log n)$, which is relevant since the round complexity of interactive MPC protocols, such as GMW, grows with depth; in total we need $O(|P| \cdot \log n)$ rounds of communication.

The LWZ protocol, on the other hand, avoids the need for general purpose MPC circuit evaluation. Indeed, the protocol is strikingly lightweight: The LWZ protocol does not require execution of any secure gates at all. The downside of LWZ is that its performance degrades with the number of *tolerated corruptions*, $t$. I.e., suppose that at most $t$ parties will collude and share information with one another. To prevent these adversaries from learning the final permutation of the elements, we must ensure that for *each* subset of $t$ parties, there exists one repetition of the protocol where *none* of those parties is on the committee. Thus, we must make our committees each of size $|P|-t$, and the number of needed repetitions grows with $\binom{|P|}{|P|-t}$. If $t$ is small, say $t=1$, then the LWZ protocol has excellent performance requiring only $|P|$ rounds of communication. If $t$ is large, say $t=\frac{|P|}{2}-1$, then performance degrades exponentially in $|P|$.

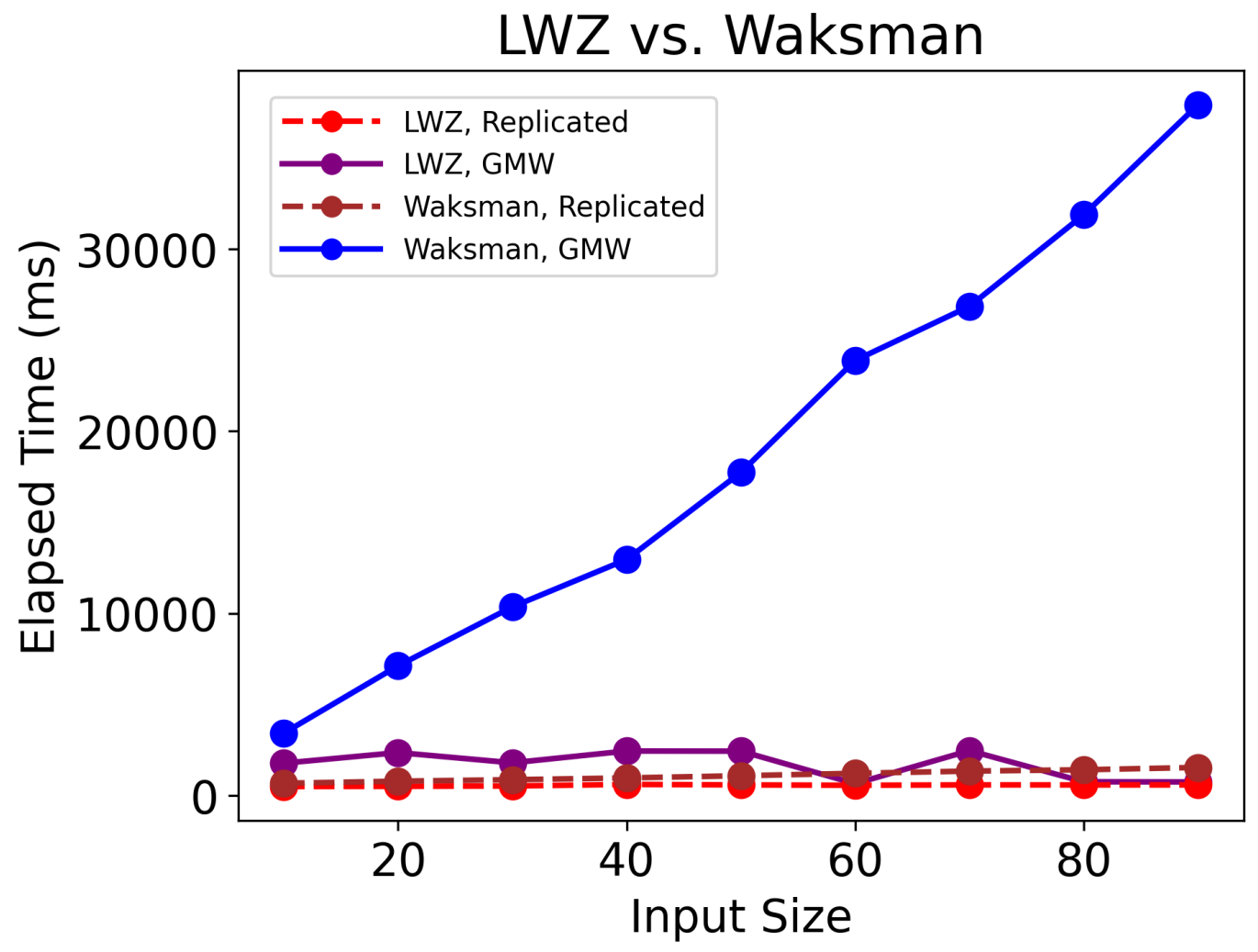

**Figure 8** End-to-end execution time of Waksman vs LWZ shuffle over three parties, averaged over five samples (lower is better). The **Replicated** protocol executes the program without cryptography, executing operations in the clear. The **GMW** protocol uses the SYMPHONY implementation of GMW which uses MOTION as a backend. **Input Size** indicates the length of the integer list provided as input by each party.

**SYMPHONY Execution Time** Figure 8 plots SYMPHONY's end-to-end execution time for the LWZ and Waksman shuffles. In both protocols, three parties each share an array of **Input Size** integers which are concatenated and shuffled. We ran the programs using both the **GMW** protocol using the **Replicated** protocol as a baseline. In the **Replicated** protocol, Boolean gates are implemented locally and computed in the clear instead of via cryptography. For our LWZ threshold, we chose the optimistic setting where the maximum number of colluding parties is $t=1$.

Our results demonstrate that the SYMPHONY implementation of LWZ properly avoids MPC overhead: as already stated, LWZ is a lightweight protocol, so SYMPHONY should not – and does not – erroneously introduce cost just because we are operating on GMW shares. As expected, we find that the **GMW**-based Waksman implementation is much slower than both **Replicated** Waksman and both variants of LWZ. The slowdown is primarily due to the cryptography required to execute the Boolean gates under MPC.

We do note that **GMW**-based Waksman achieves lower performance than might be expected. We observed that the low performance is due to the MOTION backend which, on this benchmark, allocates $> 4$ GB of memory per party to store the GMW circuit. Moreover, the execution of each gate involves accessing many non-contiguous memory addresses, leading to low spacial locality. We believe that performance can be greatly increased by handling more of the circuit generation and execution in the compatibility layer of Symphony.

Even with a highly optimized GMW backend, the LWZ protocol would remain best for the setting of $t = 1$: the coordination-heavy LWZ protocol is simply a superior technique for the setting. Symphony's features make the complex coordination involved in this protocol easy to express.

# B Metatheory

## B.1 Single-Threaded Semantics

Figure 9 shows the complete single-threaded semantics which are discussed in Section 4.2.

Notice that rules for handling I/O require that the mode is a singleton party; this is important for ensuring compatibility (i.e., so that all parties agree on the contents of shared variables).

Sums and pairs are essentially standard, modulo the consideration of their values' locations, and party sets are constructed via set-union, and deconstructed via pattern matching.

Symphony directly supports lists and arrays; in $\lambda$-Symphony they can be encoded by iterated sum and pair values where `nil` $\triangleq \iota_1\ 0$ and `cons` $\triangleq \lambda x.\ \lambda xs.\ \iota_2\ \langle x, xs\rangle$; lists can be deconstructed by pattern matching with `case`. We can encode bundles as an *association list*, implementing a map from parties to values located at that party. For example, the following list represents a bundle with 8 located at $A$ and 3 located at $B$.

$$\iota_2\ \langle\langle\{A\}, 8@\{A\}\rangle, \iota_2\ \langle\langle\{B\}, 3@\{B\}\rangle, (\iota_1\ 0)\rangle\rangle$$

(Missing location annotations for the list itself are dropped to avoid clutter; they are all $@\{A,B\}$.)

## B.2 Proof Sketches for Correspondence Theorems

To prove theorems Theorem 6.1, Theorem 6.2 and Theorem 6.3 given in Section 6, we first formalize key definitions.

**Definition B.1** (Terminal State).

$$\begin{aligned}
\varsigma \text{ is a terminal state} &\overset{\triangle}{\iff} \varsigma = m, \gamma, \delta, \top, a \wedge \gamma \vdash_m \delta, a \hookrightarrow \delta', v \\
\dot{\varsigma} \text{ is a terminal state} &\overset{\triangle}{\iff} \dot{\varsigma} = m, \dot{\gamma}, \dot{\delta}, \top, a \wedge \dot{\gamma} \vdash_m \dot{\delta}, a \hookrightarrow \dot{\delta}', \dot{v} \\
C \text{ is a terminal state} &\overset{\triangle}{\iff} \forall A \in dom(C).\ C(A) \text{ is a terminal state}
\end{aligned}$$

This definition captures the idea that a state is terminal if the execution stack is empty ($\top$), the next term to execute is atomic ($a$), and the atomic expression is able

$$\kappa \in \text{stack} ::= \top \mid \langle \texttt{let}\ x = \square\ \texttt{in}\ e \mid m, \gamma \rangle :: \kappa \qquad \varsigma \in \text{config} ::= m, \gamma, \delta, \kappa, e$$

$\boxed{\gamma \vdash_m \delta, a \hookrightarrow \delta, v}$

ST-VAR
$$\frac{}{\gamma \vdash_m \delta, x \hookrightarrow \delta, \gamma(x)\surd_m}$$

ST-LIT
$$\frac{}{\begin{array}{c}\gamma \vdash_m \delta, i \hookrightarrow \delta, i@m \\ \gamma \vdash_m \delta, p \hookrightarrow \delta, p@m\end{array}}$$

ST-INT-BINOP
$$\frac{\begin{array}{c} i_1^{\psi}@m = \gamma(x_1)\surd_m \\ i_2^{\psi}@m = \gamma(x_2)\surd_m \qquad \vdash_m \psi \end{array}}{\gamma \vdash_m \delta, x_1 \odot x_2 \hookrightarrow \delta, [\![\odot]\!](i_1, i_2)^{\psi}@m}$$

ST-PSET-BINOP
$$\frac{\begin{array}{c} p_1@m = \gamma(x_1)\surd_m \\ p_2@m = \gamma(x_2)\surd_m \end{array}}{\gamma \vdash_m \delta, x_1 \cup x_2 \hookrightarrow \delta, (p_1 \cup p_2)@m}$$

ST-MUX
$$\frac{\begin{array}{c} i_1^{\psi}@m = \gamma(x_1)\surd_m \\ i_2^{\psi}@m = \gamma(x_2)\surd_m \\ i_3^{\psi}@m = \gamma(x_3)\surd_m \qquad \vdash_m \psi \end{array}}{\gamma \vdash_m \delta, x_1\ ?\ x_2 \diamond x_3 \hookrightarrow \delta, \text{cond}(i_1, i_2, i_3)^{\psi}@m}$$

ST-PAIR
$$\frac{\begin{array}{c} v_1 = \gamma(x_1)\surd_m \\ v_2 = \gamma(x_2)\surd_m \end{array}}{\gamma \vdash_m \delta, \langle x_1, x_2 \rangle \hookrightarrow \delta, \langle v_1, v_2 \rangle@m}$$

ST-PROJ
$$\frac{\langle v_1, v_2 \rangle@m = \gamma(x)\surd_m}{\gamma \vdash_m \delta, \pi_i\ x \hookrightarrow \delta, v_i}$$

ST-INJ
$$\frac{v = \gamma(x)\surd_m}{\gamma \vdash_m \delta, \iota_i\ x \hookrightarrow \delta, (\iota_i\ v)@m}$$

ST-FUN
$$\frac{}{\gamma \vdash_m \delta, \lambda_z x.\ e \hookrightarrow \delta, \langle \lambda_z x.\ e, \gamma \rangle@m}$$

ST-REF
$$\frac{v = \gamma(x)\surd_m}{\gamma \vdash_m \delta, \texttt{ref}\ x \hookrightarrow \{\ell \mapsto v\} \uplus \delta, \ell^{\#m}@m}$$

ST-DEREF
$$\frac{\ell^{\#q}@m = \gamma(x)\surd_m}{\gamma \vdash_m \delta, !x \hookrightarrow \delta, \delta(\ell)\surd_m}$$

ST-ASSIGN
$$\frac{\begin{array}{c} \ell^{\#m}@m = \gamma(x_1)\surd_m \\ v = \gamma(x_2)\surd_m \end{array}}{\gamma \vdash_m \delta, x_1 := x_2 \hookrightarrow \delta[\ell \mapsto v], v}$$

ST-READ
$$\frac{|m| = 1}{\gamma \vdash_m \delta, \texttt{read} \hookrightarrow \delta, i@m}$$

ST-WRITE
$$\frac{i@m = \gamma(x)\surd_m \qquad |m| = 1}{\gamma \vdash_m \delta, \texttt{write}\ x \hookrightarrow \delta, 0@m}$$

ST-SHARE
$$\frac{\begin{array}{cc} p@m = \gamma(x_1)\surd_m & \vdash_p \psi \\ q@m = \gamma(x_2)\surd_m & q \neq \varnothing \\ i^{\psi}@p = \gamma(x_3)\surd_p & m = p \cup q \end{array}}{\gamma \vdash_m \delta, \texttt{share}[x_1 \to x_2]\ x_3 \hookrightarrow \delta, i^{\text{enc}\#q}@q}$$

ST-REVEAL
$$\frac{\begin{array}{cc} p@m = \gamma(x_1)\surd_m & \\ q@m = \gamma(x_2)\surd_m & q \neq \varnothing \\ i^{\text{enc}\#p}@p = \gamma(x_3)\surd_p & m = p \cup q \end{array}}{\gamma \vdash_m \delta, \texttt{reveal}[x_1 \to x_2]\ x_3 \hookrightarrow \delta, i@q}$$

$\boxed{\varsigma \longrightarrow \varsigma}$

ST-CASE-INJ
$$\frac{(\iota_i\ v)@m = \gamma(x_1)\surd_m}{m, \gamma, \delta, \kappa, \texttt{case}\ x_1\ \{x_2.e_1\}\{x_2.e_2\} \longrightarrow m, \{x_2 \mapsto v\} \uplus \gamma, \delta, \kappa, e_i}$$

ST-CASE-PSET-EMP
$$\frac{\varnothing@m = \gamma(x_1)\surd_m}{m, \gamma, \delta, \kappa, \texttt{case}\ x_1\ \{.e_1\}\{x_2 x_3.e_2\} \longrightarrow m, \gamma, \delta, \kappa, e_1}$$

ST-CASE-PSET-CONS
$$\frac{(\{A\} \uplus p)@m = \gamma(x_1)\surd_m}{m, \gamma, \delta, \kappa, \texttt{case}\ x_1\ \{.e_1\}\{x_2 x_3.e_2\} \longrightarrow m, \{x_2 \mapsto \{A\}, x_3 \mapsto p\} \uplus \gamma, \delta, \kappa, e_2}$$

ST-PAR
$$\frac{p@m = \gamma(x)\surd_m \qquad m \cap p \neq \varnothing}{m, \gamma, \delta, \kappa, \texttt{par}\ x\ e \longrightarrow m \cap p, \gamma, \delta, \kappa, e}$$

ST-PAREMPTY
$$\frac{p@m = \gamma(x)\surd_m \qquad m \cap p = \varnothing \qquad \gamma' = \{x' \mapsto \bigstar\} \uplus \gamma}{m, \gamma, \delta, \kappa, \texttt{par}\ x\ e \longrightarrow m, \gamma', \delta, \kappa, x'}$$

ST-APP
$$\frac{v_1 = \gamma(x_1)\surd_m \qquad v_2 = \gamma(x_2)\surd_m \qquad \langle \lambda_z x.\ e, \gamma' \rangle@m = v_1}{m, \gamma, \delta, \kappa, x_1\ x_2 \longrightarrow m, \{z \mapsto v_1, x \mapsto v_2\} \uplus \gamma', \delta, \kappa, e}$$

ST-LETPUSH
$$\frac{\kappa' = \langle \texttt{let}\ x = \square\ \texttt{in}\ e_2 \mid m, \gamma \rangle :: \kappa}{m, \gamma, \delta, \kappa, \texttt{let}\ x = e_1\ \texttt{in}\ e_2 \longrightarrow m, \gamma, \delta, \kappa', e_1}$$

ST-LETPOP
$$\frac{\gamma \vdash_m \delta, a \hookrightarrow \delta', v \qquad \kappa = \langle \texttt{let}\ x = \square\ \texttt{in}\ e \mid m', \gamma' \rangle :: \kappa'}{m, \gamma, \delta, \kappa, a \longrightarrow m', \{x \mapsto v\} \uplus \gamma', \delta', \kappa', e}$$

**Figure 9** $\lambda$-SYMPHONY single-threaded semantics. Premises highlighted are required only for lazy MPC evaluation.

to step (via $\hookrightarrow$) to a value $v$. There are no successor configurations which can be reached from a terminal state. Any state which is both non-terminal and also has no successor configurations we call *stuck*.

**Definition B.2** (Divergence). *A single-threaded configuration $\varsigma$ is divergent if for all $\varsigma'$ where $\varsigma \longrightarrow^* \varsigma'$, there exists $\varsigma''$ s.t. $\varsigma' \longrightarrow \varsigma''$. (And likewise for distributed configurations $C$ and transitions $\rightsquigarrow^*$.)*

**Definition B.3** (Locally stuck).

$$\begin{array}{lll} C \text{ is locally stuck} \overset{\Delta}{\iff} \exists A \text{ s.t. } C(A) = \dot{\varsigma} & & \\ \quad \textit{and where} & \dot{\varsigma} \text{ is not a terminal state} & \\ & \dot{\varsigma}.e \notin \{share[_ \rightarrow _]\ _, reveal[_ \rightarrow _]\ _\} & \\ & \dot{\varsigma} \not\longrightarrow_A & \\ \quad \textit{or} & \dot{\varsigma}.e \in \{share[x_1 \rightarrow x_2]\ x_3, reveal[x_1 \rightarrow x_2]\ x_3\} & \\ & p = \dot{\varsigma}.\dot{\gamma}(x_1) & m = \dot{\varsigma}.m \\ & q = \dot{\varsigma}.\dot{\gamma}(x_2) & m \neq p \cup q \end{array}$$

Now we establish a number of key lemmas. Our proof approach for Theorem 6.1 largely follows the proof approach from Wysteria [28], and our proof approach for Theorem 6.2 and Theorem 6.3—while novel—are straightforward proofs by case analysis and inductive reasoning on the recursive syntax of configurations and inductively defined relations $\longrightarrow$, $\rightsquigarrow$ and $\longrightarrow_A$. In this section we show the high level proof approach.

First, we establish determinism for the single-threaded semantics and confluence for the distributed semantics:

**Lemma B.1** (ST Determinism). *If $\varsigma \longrightarrow \varsigma_1$ and $\varsigma \longrightarrow \varsigma_2$ then $\varsigma_1 = \varsigma_2$.*

*Proof.* Case analysis on derivations $\varsigma \longrightarrow \varsigma_1$ and $\varsigma \longrightarrow \varsigma_2$. ☐

**Lemma B.2** (D Confluence). *If $C \rightsquigarrow^* C_1$ and $C \rightsquigarrow^* C_2$ then $C_1 \rightsquigarrow^* C_3$ and $C_2 \rightsquigarrow^* C_3$ for some $C_3$.*

*Proof.* We first prove a diamond property sublemma that shows if $C \rightsquigarrow C_1$, $C \rightsquigarrow C_2$ and $C_1 \neq C_2$, then $C_1 \rightsquigarrow C_3$ and $C_2 \rightsquigarrow C_3$ for some $C_3$, which is proved by case analysis on derivations $C \rightsquigarrow C_1$ and $C \rightsquigarrow C_2$. Confluence is established as a classic results whereby transition systems which satisfy the diamond property are also confluent, the proof of which is by induction on derivations $C \rightsquigarrow^* C_1$ and $C \rightsquigarrow^* C_2$ and appealing to the diamond property in the base cases. ☐

Next, we establish forward simulation between terminal states and semantics:

**Lemma B.3** (ST Forward Simulation).

1. *If $\varsigma$ is terminal then $\varsigma\lightning$ is terminal*
2. *If $\varsigma$ is stuck then $\varsigma\lightning$ is locally stuck*
3. *If $\varsigma \longrightarrow^* \varsigma'$ then $\varsigma\lightning \rightsquigarrow^* \varsigma'\lightning$.*

*Proof.*

1. Case analysis on $\varsigma$
2. Case analysis on $\varsigma$
3. Induction on steps in $\varsigma \longrightarrow^* \varsigma'$ and case analysis on intermediate derivations $\varsigma \longrightarrow \varsigma''$.

□

Theorem 6.1 then follows from these lemmas:

*Proof of ST/D Terminal Correspondence.* The forward direction is equivalent to showing $\varsigma \longrightarrow^* \varsigma'$ and $\varsigma'$ terminal implies $\varsigma\lightning \longrightarrow^* \varsigma'\lightning$ and $\varsigma'\lightning$ terminal, which follows from Lemma B.3.

The backward direction is equivalent to showing $\varsigma\lightning \rightsquigarrow^* C$ and $C$ terminal implies $\varsigma \longrightarrow^* \varsigma'$ for some $\varsigma'$ where $\varsigma'$ terminal and $C = \varsigma'\lightning$. By Lemma B.3 and Lemma B.2 we know that if $\varsigma$ diverges then $\varsigma\lightning$ must diverge, and therefore under the assumption that $\varsigma\lightning$ converges, we know must converge, so $\varsigma \longrightarrow^* \varsigma'$ for some terminal state $\varsigma'$. By Lemma B.3 we know $\varsigma\lightning \rightsquigarrow^* \varsigma'\lightning$, and by Lemma B.2 we know $C = \varsigma'\lightning$. □

Our proof of Theorem 6.2 also follows from the lemmas and theorem proven thus far:

*Proof of ST/D Strong Asymmetric Non-terminal Correspondence.*

1. By Theorem 6.1 we know $\varsigma$ doesn't reach a terminal state, so it either diverges or converges to a stuck state. Consider each case. Assume $\varsigma$ diverges, then we know by Lemma B.3 we know that there exists a distributed trace that also diverges. By Lemma B.2 applied to the stuck distributed state, the divergent distributed state (just established), and $\varsigma\lightning$ as the common ancestor, we know the stuck distributed state can make progress towards a divergent one, which is a contradiction—so this subcase can never happen. The other subcase is when $\varsigma$ reaches a stuck state, which trivially satisfies the goal.
2. Because $\rightsquigarrow$ is confluent by Lemma B.2, $\varsigma\lightning$ must either converge to a terminal state, converge to a stuck state, or diverge. (E.g., it impossible for $\varsigma\lightning \rightsquigarrow^* \varsigma'$ where $\varsigma'$ is stuck, and for $\varsigma\lightning \rightsquigarrow^* \varsigma''$ where $\varsigma''$ can continue to transition without ever reaching a stuck or terminal state.) If $\varsigma\lightning$ converged then by Theorem 6.1, which would reach a contradiction. If $\varsigma\lightning$ reached a stuck state, then so would $\varsigma$ by (1) of this theorem, which would reach a contradiction. Therefore, $\varsigma\lightning$ must diverge.

□

We prove one final lemma before proving our third theorem:

**Lemma B.4** (D Local Stuck Preservation). *If $C$ is locally stuck and $C \rightsquigarrow^* C'$ then $C'$ is locally stuck.*

*Proof.* Induction on the number of steps in $\rightsquigarrow^*$, and case analysis on intermediate derivations $C \rightsquigarrow C''$. □

Our proof of Theorem 6.3 then uses the prior lemma:

*Proof of ST/D Soundness for Stuck States.* We assume $\varsigma \longrightarrow^* \varsigma'$ where $\varsigma'$ is stuck and some $C$ where $\varsigma\lightning \rightsquigarrow C$. We must show there exists $C'$ s.t. $C \rightsquigarrow C'$ and $C'$ locally stuck. By Lemma B.3 we know $\varsigma\lightning \rightsquigarrow^* \varsigma'\lightning$ and $\varsigma'\lightning$ is locally stuck. By confluence we have there exists $C'$ s.t. $C \rightsquigarrow^* C'$ and $\varsigma'\lightning \rightsquigarrow^* C'$. By Lemma B.4 with $\varsigma\lightning$ as the common ancestor we have $C'$ locally stuck. □

Theorem 6.1 captures the same metatheoretical properties proved of prior work (Wysteria [28]), whereas Theorems 6.2 and 6.3 are refinements of divergence-soundness and stuck-state-soundness results novel to our work.

### B.3 Detailed Proofs for Key Lemmas

In this section, we prove the key meta-theoretic properties of the distributed semantics, namely forward simulation (Appendix B.3.1) and confluence (Appendix B.3.2), along with their corollaries. The full DS-semantics rules are given in Figure 10.

#### B.3.1 Forward Simulation

The key lemma for proving simulation states that if global single-threaded configuration $\varsigma$ steps to $\varsigma'$, then the slicing of $\varsigma$ steps to $\varsigma'$ over multiple steps of the multi-threaded semantics. The basic structure of the proof is, based on the form of step from global configuration $\varsigma$, to construct a sequence of distributed steps that each updates the local configuration of some party in the mode of $\varsigma$. For non-atomic expressions, there is exactly one step for every party in the mode; the most interesting case are global steps that are applications SS-Par: these are simulated by a sequence of steps which may be built from applications of SS-Par themselves *or* SS-Empty. For expressions that evaluate an atom and bind the result, there is a single step, performed by all parties.

**Lemma B.5** (Forward Simulation-Step)**.** *If $\varsigma \rightarrow \varsigma'$, then $\varsigma\lightning \rightsquigarrow^* \varsigma'\lightning$.*

*Proof.* $\varsigma \rightarrow \varsigma'$, by assumption. Let $(m, \gamma, \delta, \kappa, e) = \varsigma$ and let $(m', \gamma', \delta', \kappa', e') = \varsigma'$. Proceed by cases on the form of the evidence of $\varsigma \rightarrow \varsigma'$:

**ST-Case-Inj** $\varsigma\lightning \rightsquigarrow \ldots \rightsquigarrow C_i \rightsquigarrow \ldots \rightsquigarrow \varsigma'\lightning$, where each $C_i$ is $\varsigma\lightning|_{[0,i]} \uplus \varsigma'\lightning|_{[i+1,|m|]}$ (where $C|_I$ denotes distributed configuration $C$ restricted to parties at indices $I$).

The proof that each $C_i$ steps to $C_{i+1}$ is as follows. Apply DS-Step, with $\dot{\varsigma}$ as the configuration

$$m_i, \gamma, \delta, \kappa, \text{case } x\{x_1.e_1\}\{x_2.e_2\}$$

$\dot{\varsigma}'$ as the configuration

$$m_i, \{x \mapsto v\} \uplus \gamma, \delta, \kappa, e_j$$

where $\gamma(x) \downarrow_{m_i} = (\iota_j v)@\{m_i\}$ and $C_i|_{[0,i-1],[i+1,|m|]}$ as $C$. $\dot{\varsigma} \longrightarrow_i \dot{\varsigma}'$ by DS-Case-Inj. $C_{i+1}$ is $C_i|_{[0,i-1]} \uplus \{m_i \mapsto \dot{\varsigma}'\} \uplus C_i|_{[i+1,|m|]}$.

The proofs for evidence constructed from rules DS-Case-PSet-Emp and DS-Case-PSet-Cons are similar. The only distinction is that the updated local configuration in each distributed configuration $C_{i+1}$ is formed by updating the subject of expression

$$\dot{v} \in \text{lval} ::= i^{\psi} \mid p \mid \ell^{\#m} \mid \iota_i\ \dot{v} \mid \langle \dot{v}, \dot{v} \rangle \mid \langle \lambda_z x.\ e, \dot{\gamma} \rangle \mid \bigstar$$

$$\dot{\gamma} \in \text{lenv} \triangleq \text{var} \rightharpoonup \text{lval} \qquad \dot{\delta} \in \text{lstore} \triangleq \text{loc} \rightharpoonup \text{lval} \qquad \dot{\kappa} \in \text{lstack} ::= \top \mid \langle \texttt{let}\ x = \square\ \texttt{in}\ e \mid m, \dot{\gamma} \rangle :: \dot{\kappa}$$

$$\dot{\varsigma} \in \text{lconfig} ::= m, \dot{\gamma}, \dot{\delta}, \dot{\kappa}, e \qquad C \in \text{dconfig} \triangleq \text{party} \rightharpoonup \text{lconfig}$$

$$\boxed{\dot{\gamma} \vdash_m \dot{\delta}, a \hookrightarrow \dot{\delta}, \dot{v}}$$

DS-VAR
$$\frac{}{\dot{\gamma} \vdash_m \dot{\delta}, x \hookrightarrow \dot{\delta}, \dot{\gamma}(x)}$$

DS-LIT
$$\frac{}{\dot{\gamma} \vdash_m \dot{\delta}, i \hookrightarrow \dot{\delta}, i \qquad \dot{\gamma} \vdash_m \dot{\delta}, p \hookrightarrow \dot{\delta}, p}$$

DS-INT-BINOP
$$\frac{i_1^{\psi} = \dot{\gamma}(x_1) \quad i_2^{\psi} = \dot{\gamma}(x_2) \quad \vdash_m \psi}{\dot{\gamma} \vdash_m \dot{\delta}, x_1 \odot x_2 \hookrightarrow \dot{\delta}, [\![\odot]\!](i_1, i_2)^{\psi}}$$

DS-PSET-BINOP
$$\frac{p_1 = \dot{\gamma}(x_1) \quad p_2 = \dot{\gamma}(x_2)}{\dot{\gamma} \vdash_m \dot{\delta}, x_1 \cup x_2 \hookrightarrow \dot{\delta}, p_1 \cup p_2}$$

DS-MUX
$$\frac{i_1^{\psi} = \dot{\gamma}(x_1) \quad i_2^{\psi} = \dot{\gamma}(x_2) \quad i_3^{\psi} = \dot{\gamma}(x_3) \quad \vdash_m \psi}{\dot{\gamma} \vdash_m \dot{\delta}, x_1\ ?\ x_2 \diamond x_3 \hookrightarrow \dot{\delta}, \text{cond}(i_1, i_2, i_3)^{\psi}}$$

DS-PAIR
$$\frac{\dot{v}_1 = \dot{\gamma}(x_1) \quad \dot{v}_2 = \dot{\gamma}(x_2)}{\dot{\gamma} \vdash_m \dot{\delta}, \langle x_1, x_2 \rangle \hookrightarrow \dot{\delta}, \langle \dot{v}_1, \dot{v}_2 \rangle}$$

DS-PROJ
$$\frac{\langle \dot{v}_1, \dot{v}_2 \rangle = \dot{\gamma}(x)}{\dot{\gamma} \vdash_m \dot{\delta}, \pi_i\ x \hookrightarrow \dot{\delta}, \dot{v}_i}$$

DS-INJ
$$\frac{\dot{v} = \dot{\gamma}(x)}{\dot{\gamma} \vdash_m \dot{\delta}, \iota_i\ x \hookrightarrow \dot{\delta}, (\iota_i\ \dot{v})}$$

DS-FUN
$$\frac{}{\dot{\gamma} \vdash_m \dot{\delta}, \lambda_z x.\ e \hookrightarrow \dot{\delta}, \langle \lambda_z x.\ e, \dot{\gamma} \rangle}$$

DS-REF
$$\frac{\dot{v} = \dot{\gamma}(x)}{\dot{\gamma} \vdash_m \dot{\delta}, \texttt{ref}\ x \hookrightarrow \{\ell \mapsto \dot{v}\} \uplus \dot{\delta}, \ell^{\#m}}$$

DS-DEREF
$$\frac{\ell^{\#q} = \dot{\gamma}(x)}{\dot{\gamma} \vdash_m \dot{\delta}, !x \hookrightarrow \dot{\delta}, \dot{\delta}(\ell)}$$

DS-ASSIGN
$$\frac{\ell^{\#m} = \dot{\gamma}(x_1) \quad \dot{v} = \dot{\gamma}(x_2)}{\dot{\gamma} \vdash_m \dot{\delta}, x_1 := x_2 \hookrightarrow \dot{\delta}[\ell \mapsto \dot{v}], \dot{v}}$$

DS-READ
$$\frac{|m| = 1}{\dot{\gamma} \vdash_m \dot{\delta}, \texttt{read} \hookrightarrow \dot{\delta}, i}$$

DS-WRITE
$$\frac{i = \dot{\gamma}(x) \quad |m| = 1}{\dot{\gamma} \vdash_m \dot{\delta}, \texttt{write}\ x \hookrightarrow \dot{\delta}, 0}$$

$$\boxed{\dot{\varsigma} \longrightarrow_A \dot{\varsigma}}$$

DS-CASE-INJ
$$\frac{(\iota_i\ \dot{v}) = \dot{\gamma}(x_1)}{m, \dot{\gamma}, \dot{\delta}, \dot{\kappa}, \texttt{case}\ x_1\ \{x_2.e_1\}\{x_2.e_2\} \longrightarrow_A m, \{x_2 \mapsto \dot{v}\} \uplus \dot{\gamma}, \dot{\delta}, \dot{\kappa}, e_i}$$

DS-CASE-PSET-EMP
$$\frac{\varnothing = \dot{\gamma}(x_1)}{m, \dot{\gamma}, \dot{\delta}, \dot{\kappa}, \texttt{case}\ x_1\ \{.e_1\}\{x_2 x_3.e_2\} \longrightarrow_A m, \dot{\gamma}, \dot{\delta}, \dot{\kappa}, e_1}$$

DS-CASE-PSET-CONS
$$\frac{(\{B\} \uplus p) = \dot{\gamma}(x_1)}{m, \dot{\gamma}, \dot{\delta}, \dot{\kappa}, \texttt{case}\ x_1\ \{.e_1\}\{x_2 x_3.e_2\} \longrightarrow_A m, \{x_2 \mapsto \{B\}, x_3 \mapsto p\} \uplus \dot{\gamma}, \dot{\delta}, \dot{\kappa}, e_2}$$

DS-PAR
$$\frac{p = \dot{\gamma}(x) \quad A \in p}{m, \dot{\gamma}, \dot{\delta}, \dot{\kappa}, \texttt{par}\ x\ e \longrightarrow_A m \cap p, \dot{\gamma}, \dot{\delta}, \dot{\kappa}, e}$$

DS-PAREMPTY
$$\frac{p = \dot{\gamma}(x) \quad A \notin p \quad \dot{\gamma}' = \{x' \mapsto \bigstar\} \uplus \dot{\gamma}}{m, \dot{\gamma}, \dot{\delta}, \dot{\kappa}, \texttt{par}\ x\ e \longrightarrow_A m, \dot{\gamma}', \dot{\delta}, \dot{\kappa}, x'}$$

DS-APP
$$\frac{\dot{v}_1 = \dot{\gamma}(x_1) \quad \dot{v}_2 = \dot{\gamma}(x_2) \quad \langle \lambda_z x.\ e, \dot{\gamma}' \rangle = \dot{v}_1}{m, \dot{\gamma}, \dot{\delta}, \dot{\kappa}, x_1\ x_2 \longrightarrow_A m, \{z \mapsto \dot{v}_1, x \mapsto \dot{v}_2\} \uplus \dot{\gamma}', \dot{\delta}, \dot{\kappa}, e}$$

DS-LETPUSH
$$\frac{\dot{\kappa}' = \langle \texttt{let}\ x = \square\ \texttt{in}\ e_2 \mid m, \dot{\gamma} \rangle :: \dot{\kappa}}{m, \dot{\gamma}, \dot{\delta}, \dot{\kappa}, \texttt{let}\ x = e_1\ \texttt{in}\ e_2 \longrightarrow_A m, \dot{\gamma}, \dot{\delta}, \dot{\kappa}', e_1}$$

DS-LETPOP
$$\frac{\dot{\gamma} \vdash_m \dot{\delta}, a \hookrightarrow \dot{\delta}', \dot{v} \quad \dot{\kappa} = \langle \texttt{let}\ x = \square\ \texttt{in}\ e \mid m', \dot{\gamma}' \rangle :: \dot{\kappa}'}{m, \dot{\gamma}, \dot{\delta}, \dot{\kappa}, a \longrightarrow_A m', \{x \mapsto \dot{v}\} \uplus \dot{\gamma}', \dot{\delta}', \dot{\kappa}', e}$$

$$\boxed{C \rightsquigarrow C}$$

DS-STEP
$$\frac{\dot{\varsigma} \longrightarrow_A \dot{\varsigma}'}{\{A \mapsto \dot{\varsigma}\} \uplus C \rightsquigarrow \{A \mapsto \dot{\varsigma}'\} \uplus C}$$

DS-SHARE
$$\frac{\begin{array}{lll} \texttt{share}[x_1 \to x_2]\ x_3 = C(m).e & \vdash_p \psi & C' = \{A \mapsto (m, \{x \mapsto \dot{v}\} \uplus \dot{\gamma}, \dot{\delta}, \dot{\kappa}, x) \\ p = C(m).\dot{\gamma}(x_1) & m = C(m).m & \mid C(A) = (m, \dot{\gamma}, \dot{\delta}, \dot{\kappa}, e), \\ q = C(m).\dot{\gamma}(x_2) & m = p \cup q & A \in q \implies \dot{v} = i^{\text{enc}\#q}, \\ i^{\psi} = C(p).\dot{\gamma}(x_3) & q \neq \varnothing & A \in p \land A \notin q \implies \dot{v} = \bigstar\} \end{array}}{C_0 \uplus C \rightsquigarrow C_0 \uplus C'}$$

DS-REVEAL
$$\frac{\begin{array}{lll} \texttt{reveal}[x_1 \to x_2]\ x_3 = C(m).e & & C' = \{A \mapsto (m, \{x \mapsto \dot{v}\} \uplus \dot{\gamma}, \dot{\delta}, \dot{\kappa}, x) \\ p = C(m).\dot{\gamma}(x_1) & m = C(m).m & \mid C(A) = (m, \dot{\gamma}, \dot{\delta}, \dot{\kappa}, e), \\ q = C(m).\dot{\gamma}(x_2) & m = p \cup q & A \in q \implies \dot{v} = i, \\ i^{\text{enc}\#p} = C(p).\dot{\gamma}(x_3) & q \neq \varnothing & A \in p \land A \notin q \implies \dot{v} = \bigstar\} \end{array}}{C_0 \uplus C \rightsquigarrow C_0 \uplus C'}$$

**Figure 10** λ-SYMPHONY distributed semantics (full figure).

to $e_1$ in the case of Rule DS-Case-PSet-Emp and $e_2$ in the case of Rule DS-Case-PSet-Cons. Additionally, in the case of Rule DS-Case-PSet-Cons, the local state is updated to bind variables $x_2$ and $x_3$ to the deconstructed principal and remaining set of principals.

**ST-Par** $\varsigma\lightning \rightsquigarrow \ldots \rightsquigarrow C_i \rightsquigarrow \ldots \rightsquigarrow \varsigma'\lightning$, where each $C_i$ is $\varsigma\lightning|_{[0,i]} \uplus \varsigma'\lightning|_{[i+1,|m|]}$.
The proof that each $C_i$ steps to $C_{i+1}$ is as follows. If $m_i \in p$, then apply DS-Step, with $\dot\varsigma$ as the configuration

$$m_i, \gamma, \delta, \kappa, \texttt{par}\ p\ e$$

$\dot\varsigma'$ as the configuration

$$m_i, \gamma, \delta, \kappa, e$$

and $C_i|_{[0,i-1],[i+1,|m|]}$ as $C$. $\dot\varsigma \longrightarrow_i \dot\varsigma'$ by DS-Par, because $m_i \in p$ and thus $\{m_i\} \cap p = \{m_i\} \neq \emptyset$.
If $m_i \notin p$, then let $\dot\varsigma' = \dot\varsigma$.
In both cases, $C_{i+1}$ is $C_i|_{[0,i-1]} \uplus \{m_i \mapsto \dot\varsigma'\} \uplus C_i|_{[i+1,|m|]}$.

**ST-ParEmpty** $\varsigma\lightning \rightsquigarrow \ldots \rightsquigarrow C_i \rightsquigarrow \ldots \rightsquigarrow \varsigma'\lightning$, where each $C_i$ is $\varsigma\lightning|_{[0,i]} \uplus \varsigma'\lightning|_{[i+1,|m|]}$.
The proof that each $C_i$ steps to $C_{i+1}$ is as follows. Apply DS-Step, with $\dot\varsigma$ as the configuration

$$m_i, \gamma, \delta, \kappa, \texttt{par}\ p\ e$$

$\dot\varsigma'$ as the configuration

$$m_i, \{x \mapsto \bigstar\} \uplus \gamma, \delta, \kappa, x$$

and $C_i|_{[0,i-1],[i+1,|m|]}$ as $C$. $\dot\varsigma \longrightarrow_i \dot\varsigma'$ by DS-ParEmpty, because $m \cap p = \emptyset$ by the fact that $c \rightarrow c$ is an application of ST-ParEmpty; thus $\{m_i\} \cap p = \emptyset$. $C_{i+1}$ is $C_i|_{[0,i-1]} \uplus \{m_i \mapsto \dot\varsigma'\} \uplus C_i|_{[i+1,|m|]}$.

**ST-App** $\varsigma\lightning \rightsquigarrow \ldots \rightsquigarrow C_i \rightsquigarrow \ldots \rightsquigarrow \varsigma'\lightning$, where each $C_i$ is $\varsigma\lightning|_{[0,i]} \uplus \varsigma'\lightning|_{[i+1,|m|]}$.
The proof that each $C_i$ steps to $C_{i+1}$ is as follows. Let $\langle \lambda_z x.e', \gamma' \rangle @m = v_1 = \gamma(x_1)$, which holds in the case that $c \rightarrow c'$ is an application of ST-App.
Apply DS-Step, with $\dot\varsigma$ as the configuration

$$m_i, \gamma, \delta, \kappa, x_1\ x_2$$

$\dot\varsigma'$ as the configuration

$$m_i, \{z \mapsto v_1, x_2 \mapsto \gamma(x_2)\} \uplus \gamma, \delta, \langle \texttt{let}\ x\ = _\ \texttt{in}\ e_2 \mid \gamma \rangle :: \kappa, e'$$

and $C_i|_{[0,i-1],[i+1,|m|]}$ as $C$. $\dot\varsigma \longrightarrow_i \dot\varsigma'$ by DS-App. $C_{i+1}$ is $C_i|_{[0,i-1]} \uplus \{m_i \mapsto \dot\varsigma'\} \uplus C_i|_{[i+1,|m|]}$.

**ST-LetPush** $\varsigma\lightning \rightsquigarrow \ldots \rightsquigarrow C_i \rightsquigarrow \ldots \rightsquigarrow \varsigma'\lightning$, where each $C_i$ is $\varsigma\lightning|_{[0,i]} \uplus \varsigma'\lightning|_{[i+1,|m|]}$.
The proof that each $C_i$ steps to $C_{i+1}$ is as follows. Apply DS-Step, with $\dot\varsigma$ as the configuration

$$m_i, \gamma, \delta, \kappa, \texttt{let}\ x\ =\ e_1\ \texttt{in}\ e_2$$

$\dot{\varsigma}'$ as the configuration

$$m_i, \gamma, \delta, \langle \texttt{let}\ x\ = _\ \texttt{in}\ e_2 \mid \rangle :: \kappa, e_1$$

and $C_i|_{[0,i-1],[i+1,|m|]}$ as $C$. $\dot{\varsigma} \longrightarrow_i \dot{\varsigma}'$ by DS-LetPush. $C_{i+1}$ is $C_i|_{[0,i-1]} \uplus \{m_i \mapsto \dot{\varsigma}'\} \uplus C_i|_{[i+1,|m|]}$.

**ST-LetPop** $e$ is some atom $a$, $\delta$ and $a$ step to $\delta'$ and some value $v$ under $\gamma$ in mode $m$, and $\kappa = \langle \texttt{let}\ x\ = _\ \texttt{in}\ e' \mid m', \gamma'' \rangle :: \kappa'$, and $\gamma' = \{x \mapsto v\} \uplus \gamma''$, by assumption. Proceed by cases on the fact that $\delta$ and $a$ step to $\delta'$ and some value $v$ under $\gamma$ in mode $m$. **Subcase: solo atom** In the case that the evaluation is an application of ST-Int, ST-Var, ST-Fun, ST-Inj, ST-Pair, ST-Proj, ST-Ref, ST-Deref, ST-Assign, ST-Fold, ST-Unfold, ST-Read, ST-Write, ST-Embed, and ST-Star, $\varsigma\lightning \rightsquigarrow \ldots \rightsquigarrow C_i \rightsquigarrow \ldots \rightsquigarrow \varsigma'\lightning$, where each $C_i$ is $\varsigma\lightning|_{[0,i]} \uplus \varsigma'\lightning|_{[i+1,|m|]}$. The proof that each $C_i$ steps to $C_{i+1}$ is as follows. Apply DS-Step, with $\dot{\varsigma}$ as the configuration

$$m_i, \gamma, \delta, \langle \texttt{let}\ x\ = _\ \texttt{in}\ e' \mid m', \gamma'' \rangle :: \kappa', a$$

$\dot{\varsigma}'$ as the configuration

$$m_i, \{x \mapsto v\} \uplus \gamma, \delta', \kappa', e'$$

and $C_i|_{[0,i-1],[i+1,|m|]}$ as $C$. $\dot{\varsigma} \longrightarrow_i \dot{\varsigma}'$ by DS-LetPop. $C_{i+1}$ is $C_i|_{[0,i-1]} \uplus \{m_i \mapsto \dot{\varsigma}'\} \uplus C_i|_{[i+1,|m|]}$.

**Subcases: binary operation over clear data** The subcases in which evaluation is an application of ST-Binop, $a$ is of the form $x_1\ \oplus\ x_2$, $\ddot{i}_1@m = \gamma(x_1) \not\downarrow_m$, and $\ddot{i}_2@m = \gamma(x_1) \not\downarrow_m$ or in which evaluation is an application of ST-PSet-Binop (i.e., the computation is an binary operation over clear data) is directly similar to the previous subcase.

**Subcase: mux on clear data** The subcase in which evaluation is an application of ST-Mux, $a$ is of the form $\texttt{mux if}\ x_1\ \texttt{then}\ x_2\ \texttt{else}\ x_3$, $\ddot{i}_1@m = \gamma(x_1) \not\downarrow_m$, $\ddot{i}_2@m = \gamma(x_2) \not\downarrow_m$, and $\ddot{i}_3@m = \gamma(x_3) \not\downarrow_m$ is directly similar to the previous subcases.

**Subcase: binary operation on encrypted data** For the subcase in which evaluation is an application of ST-Binop, $a$ is of the form $x_1\ \oplus\ x_2$, $i_1^{\mathsf{enc}\#m}@m = \gamma(x_1) \not\downarrow_m$, and $i_2^{\mathsf{enc}\#m}@m = \gamma(x_1) \not\downarrow_m$ (i.e., the computation is an binary operation over encrypted data), $\varsigma\lightning$ steps to $\varsigma'\lightning$ by application of DS-Step, with

$$m, \gamma, \delta, \kappa, x_1\ \oplus\ x_2$$

as $\dot{\varsigma}$,

$$m, \{x \mapsto v\} \uplus \gamma, \delta, \kappa, x$$

as $\dot{\varsigma}'$, and $\varsigma\lightning|_{\mathsf{parties}\setminus m}$ as $C$. $\dot{\varsigma}$ steps to $\dot{\varsigma}'$ by ST-Binop.

**Subcase: mux on encrypted data** The subcase in which evaluation is an application of ST-Mux, $a$ is of the form $\texttt{mux if}\ x_1\ \texttt{then}\ x_2\ \texttt{else}\ x_3$, $i_1^{\mathsf{enc}\#m}@m = \gamma(x_1) \not\downarrow_m$, $i_2^{\mathsf{enc}\#m}@m = \gamma(x_2) \not\downarrow_m$, and $i_3^{\mathsf{enc}\#m}@m = \gamma(x_3) \not\downarrow_m$ is directly similar to the previous subcase.

**Subcases: synchronization** The subcases in which evaluation is an application of ST-Share or ST-Reveal are directly similar to the previous two subcaes, in that they are simulated by a single step of the distributed semantics.

□

The proof of weak forward simulation follows directly from Lemma B.5.

**Lemma B.6** (ST Weak Forward Simulation). *If $\varsigma \longrightarrow^* \varsigma'$ and $\varsigma'$ is terminal, then $\varsigma\lightning \rightsquigarrow^* \varsigma'\lightning$ and $\varsigma'\lightning \not\rightsquigarrow$.*

*Proof.* The claim holds by induction on the multistep judgment $\varsigma \longrightarrow^* \varsigma$.

**Empty** If the trace is empty, then $\dot{\varsigma}$ is $\dot{\varsigma}'$. $\varsigma\lightning$ multi-steps to $\varsigma\lightning$ over the empty sequence of steps.

**Non-empty** If the trace is of the form $\varsigma \to \varsigma'' \to^* \varsigma'$, then $\varsigma\lightning \rightsquigarrow^* \varsigma''$ by Lemma B.5 and $\varsigma''\lightning \rightsquigarrow^* \varsigma'\lightning$ by the inductive hypothesis. $\varsigma\lightning \rightsquigarrow^* \varsigma\lightning$ by the fact that the concatenation of two traces is a trace.

□

#### B.3.2 Confluence and End-State Determinism

In order to prove the Diamond Property, we will first claim and prove a lemma that establishes that distinct sub-configurations that can step within each step of a distributed configuration in fact update the local configurations of disjoint sets of parties.

**Lemma B.7.** *For all distributed configurations $C$, $C_0$, and $C_1$ and all non-halting distributed configurations $C_0'$ and $C_1'$ such that*

$$C = C_0' \uplus C_0 = C_1' \uplus C_1$$

*one of the following cases holds:*

1. *$C_0' = C_1'$ and $C_0 = C_1$;*
2. *the domains of $C_0'$ and $C_1'$ are disjoint.*

*Proof.* Proceed by cases on whether the domains of $C_0'$ and $C_1'$ are disjoint. If so, then the second clause of the claim is satisfied.

Otherwise, there is some party $m_i$ in the domains of both $C_0'$ and $C_1'$. The domains of $C_0'$ and $C_1'$ are the same, by cases on the active expression $e_i$ in the local configuration located at $m_i$: if $e_i$ is a non-atom, a variable occurrence, an integer literal, a binary operation over integers, a binary operation over sets of principals, a multiplex, a pair creation, a pair projection, a sum injection, a function creation, a reference creation, a dereference, a reference assignment, a recursive type introduction, a read, or a write then the domains are singletons. Thus the domains are the same, because they are singletons that overlap.

In the case that the expression shares a value from $p$ to $q$, the steps from $C_0'$ and $C_1'$ are applications of DS-Share, which has a premise that the mode is $p \cup q$; thus, the domains of $C_0'$ and $C_1'$ are the identical set of parties $p \cup q$.

In the case that the expression reveals the value bound to variable $x$ to parties $q$, the steps from $C_0'$ and $C_1'$ are applications of DS-Reveal, which has a premise that the value bound to $x$ is encrypted for parties $p$, and that the active parties are $p \cup q$; thus, the domains of $C_0'$ and $C_1'$ are the same set of parties $p \cup q$. $C_0$ and $C_1$ are thus the same, given they are the restrictions of $C$ to the complements of the domains of $C'_0$ and $C'_1$, respectively. □

Using Lemma B.7, we can prove the Diamond Property for the transition relation over multi-threaded configurations.

*Proof.* There are distributed configurations $C_{0,0}$, $C_{0,1}$, $C_{1,0}$, and $C_{1,1}$ such that

$$C_{0,0} \uplus C_{0,1} = C = C_{1,0} \uplus C_{1,1}$$

and

$$\begin{aligned} C_0 &= C_{0,0} \uplus C'_{0,1} \\ C_1 &= C_{1,0} \uplus C'_{1,1} \end{aligned}$$

with $C_{0,1} \rightsquigarrow C'_{0,1}$ and $C_{1,1} \rightsquigarrow C'_{1,1}$, by inverting the facts that $C$ steps to $C_0$ and $C$ steps to $C_1$. Proceed by cases on the application of Lemma B.7 to $C$, $C_{0,0}$, $C_{0,1}$, $C_{1,0}$, and $C_{1,1}$:

**Identical** It follows immediately that $C_{0,0} = C_{1,0}$. Furthermore, it follows from a direct analysis of the multi-threaded transition relation that $C'_{0,1} = C'_{1,1}$. Thus

$$C_0 = C_{0,0} \uplus C'_{0,1} = C_{1,0} \uplus C'_{1,1} = C_1$$

by congruence. Thus for $C' = C'_0 = C'_1$, both $C \rightsquigarrow C'_0$ and $C \rightsquigarrow C'_1$.

**Disjoint** Let $C''$ be $C$ restricted to parties in $C_{0,0}$ and $C_{1,0}$ and let

$$C' = C'' \uplus C'_{0,1} \uplus C'_{1,1}$$

$C'$ is well-defined because the domains of $C'_{0,1}$ and $C'_{1,0}$, are the domains of $C_{0,1}$ and $C_{1,1}$, which are disjoint by assumption of this clause.
$C_0 \rightsquigarrow C'$ by cases on the fact that $C_0 \rightsquigarrow C'_0$: in each case, adjust the evidence to use $C'' \uplus C_1$ as the distributed configuration that remains unchanged and is joined with $C_0$. $C_1 \rightsquigarrow C'$ by a symmetric argument.

□

Given that the distributed semantics satisfies the diamond property, confluence (Lemma B.8) is a direct consequence of fundamental properties of general transition and rewrite systems.

**Lemma B.8** (DS Multi-step Confluence). *If $C \rightsquigarrow^* C_1$ and $C \rightsquigarrow^* C_2$ then there exists $C_3$ s.t. $C_1 \rightsquigarrow^* C_3$ and $C_2 \rightsquigarrow^* C_3$.*

*Proof.* Apply the fact that any binary relation that satisfies the Diamond property satisfies confluence [3] to the Diamond Property for the distributed step relation. □

An direct corollary of confluence is that all halting states reached from the same state are the same.

**Corollary B.8.1** (DS End-state Determinism). *If $C \rightsquigarrow^* C_1$ and $C \rightsquigarrow^* C_2$, $C_1 \not\rightsquigarrow$ and $C_2 \not\rightsquigarrow$ then $C_1 = C_2$.*

*Proof.* There is some distributed configuration $C'$ such that $C_1 \rightsquigarrow^* C'$ and $C_2 \rightsquigarrow^* C'$, by applying Lemma B.8 to the fact that $C \rightsquigarrow^* C_1$ and $C \rightsquigarrow^* C_2$. $C'$ is $C_1$ by the fact that $C_1$ is halting and thus $C_1$ multi-steps to $C'$ over the empty sequence of steps; $C'$ is $C_2$ by a symmetric argument. Thus, $C_1$ is $C_2$. □

## About the authors

**Ian Sweet** isweet@galois.com

**David Darais** darais@galois.com

**David Heath** heath.davidanthony@gatech.edu

**William Harris** wharris@galois.com

**Ryan Estes** restes@uvm.edu

**Michael Hicks** Michael Hicks (mwh@cs.umd.edu) now works at Amazon. This work was carried out while employed by the University of Maryland, College Park.